\documentclass[final,3p,times]{elsarticle}

\makeatletter
\def\ps@pprintTitle{%
 \let\@oddhead\@empty
 \let\@evenhead\@empty
 \def\@oddfoot{}%
 \let\@evenfoot\@oddfoot}
\makeatother

\usepackage{graphicx}
\usepackage{amssymb}
\usepackage{amsthm}
\usepackage{gensymb}
\usepackage{tabulary}
\usepackage{multirow}
\usepackage{hyperref}
\usepackage{array}
\usepackage{tablefootnote}
\usepackage{tabularx}
\usepackage{verbatim}
\usepackage{url}

\usepackage{fancyhdr}
\usepackage{lineno}

\journal{Energy Policy}

\usepackage[pdftex,dvipsnames]{xcolor}
\usepackage{soul}

\begin{document}

\begin{frontmatter}

%% Title, authors and addresses
%%TC:ignore
\title{Weathering Adaptation: Grid Infrastructure Planning in a Changing Climate}
% \subtitle{Preprint}
%%TC:endignore

%% use the tnoteref command within \title for footnotes;
%% use the tnotetext command for the associated footnote;
%% use the fnref command within \author or \address for footnotes;
%% use the fntext command for the associated footnote;
%% use the corref command within \author for corresponding author footnotes;
%% use the cortext command for the associated footnote;
%% use the ead command for the email address,
%% and the form \ead[url] for the home page:
%%
%% \title{Title\tnoteref{label1}}
%% \tnotetext[label1]{}
%% \author{Name\corref{cor1}\fnref{label2}}
%% \ead{email address}
%% \ead[url]{home page}
%% \fntext[label2]{}
%% \cortext[cor1]{}
%% \address{Address\fnref{label3}}
%% \fntext[label3]{}

%% use optional labels to link authors explicitly to addresses:
%% \author[label1,label2]{<author name>}
%% \address[label1]{<address>}
%% \address[label2]{<address>}
%%TC:ignore
\author[1,2]{Anna M. Brockway\footnotemark[1]}
\author[3]{Laurel N. Dunn\footnote{Both authors contributed equally to this work.}}
%\author[4]{Steve Weissman}
\address[1]{Energy and Resources Group} 
\address[2]{Department of Electrical Engineering and Computer Sciences} 
\address[3]{Department of Civil Engineering 
\address[4]{Both authors contributed equally to this work.} \\\vspace{5pt}December 3, 2019} %\address[4]{Goldman School of Public Policy}
\address{University of California, Berkeley}
%%TC:endignore
%%TC:ignore
\begin{abstract}
%% Text of abstract
Decisions related to electric power systems planning and operations rely on assumptions and insights informed by historic weather data and records of past performance.
Evolving climate trends are, however, changing the energy use patterns and operating conditions of grid assets, thus altering the nature and severity of risks the system faces.
Because grid assets remain in operation for decades, planning for evolving risks will require incorporating climate projections into grid infrastructure planning processes.
The current work traces a pathway for climate-aware decision-making in the electricity sector. 
We evaluate the suitability of using existing climate models and data for electricity planning and discuss their limitations.
We review the interactions between grid infrastructure and climate by synthesizing what is known about how changing environmental operating conditions would impact infrastructure utilization, constraints, and performance.
We contextualize our findings by presenting a case study of California, examining if and where climate data can be integrated into infrastructure planning processes.
The core contribution of the work is a series of nine recommendations detailing advancements in climate projections, grid modeling architecture, and disaster preparedness that would be needed to ensure that infrastructure planning decisions are robust to uncertainty and risks associated with evolving climate conditions.
\end{abstract}

\begin{keyword} %% keywords here, in the form: keyword \sep keyword
Grid infrastructure \sep Climate change \sep Uncertainty analysis \sep Infrastructure planning \sep Risk mitigation
%% MSC codes here, in the form: \MSC code \sep code
%% or \MSC[2008] code \sep code (2000 is the default)
\end{keyword}
%%TC:endignore

\end{frontmatter}

%% Start line numbering here if you want
% \linenumbers

%% main text

% INTRODUCTION
\section{Introduction and Motivation}
\label{sec:introduction}

A steady supply of electricity is fundamental to the normal and productive functioning of modern society. 
Climate change and severe weather make it more difficult to operate the electric system reliably.  
As a result, events such as Hurricanes Sandy and Maria, and recent wildfires in California, have led to blackouts. 
Electric power systems will need to adapt to new climate realities; to do so, it will be necessary to revise the models and types of data that inform operational and planning decisions \cite{Zamuda2016}. 

Decades of scientific research inform our current understanding of climate science, energy systems, and the interactions between them.
Yet questions remain about the underlying physical processes in both disciplines, as well as about how emissions will unfold over the next century.
For example, methods for interpreting global climate projections to anticipate severe weather events are still under development \cite{barsugli2013}.
The characteristics of severe weather events are also evolving, indicating that historic data are not representative of present or future conditions \cite{kunkel2013}.
Continuing to use historic data is problematic because grid infrastructure components installed today will remain in operation for decades to come.
Using climate projections to inform critical infrastructure investments may reduce our exposure to the risks we can anticipate, in light of what we currently know about climate change.

Despite abundant research characterizing climate impacts on grid infrastructure, making decisions about if and how to mitigate these impacts remains a challenge.
%We do not know precisely what the future will hold, and the science is continuously advancing.
%Using climate projections to inform long-term infrastructure investments poses a challenge to decision-makers.
It would be cost-prohibitive (and likely unnecessary) to build a system that could operate reliably in any possible climate future \cite{hallegatte2009}, and we may need to accept certain risks that we could opt to mitigate today.
Yet research shows that climate trends will fundamentally transform the performance and risk exposure of grid assets \cite{Craig2018, Bartos2016}.
Failure to incorporate climate impacts into planning decisions could leave critical infrastructure unnecessarily exposed to risks that we could feasibly avoid \cite{Biesbroek2013}.

\begin{figure}[t]
\centering
\includegraphics[width=1\linewidth]{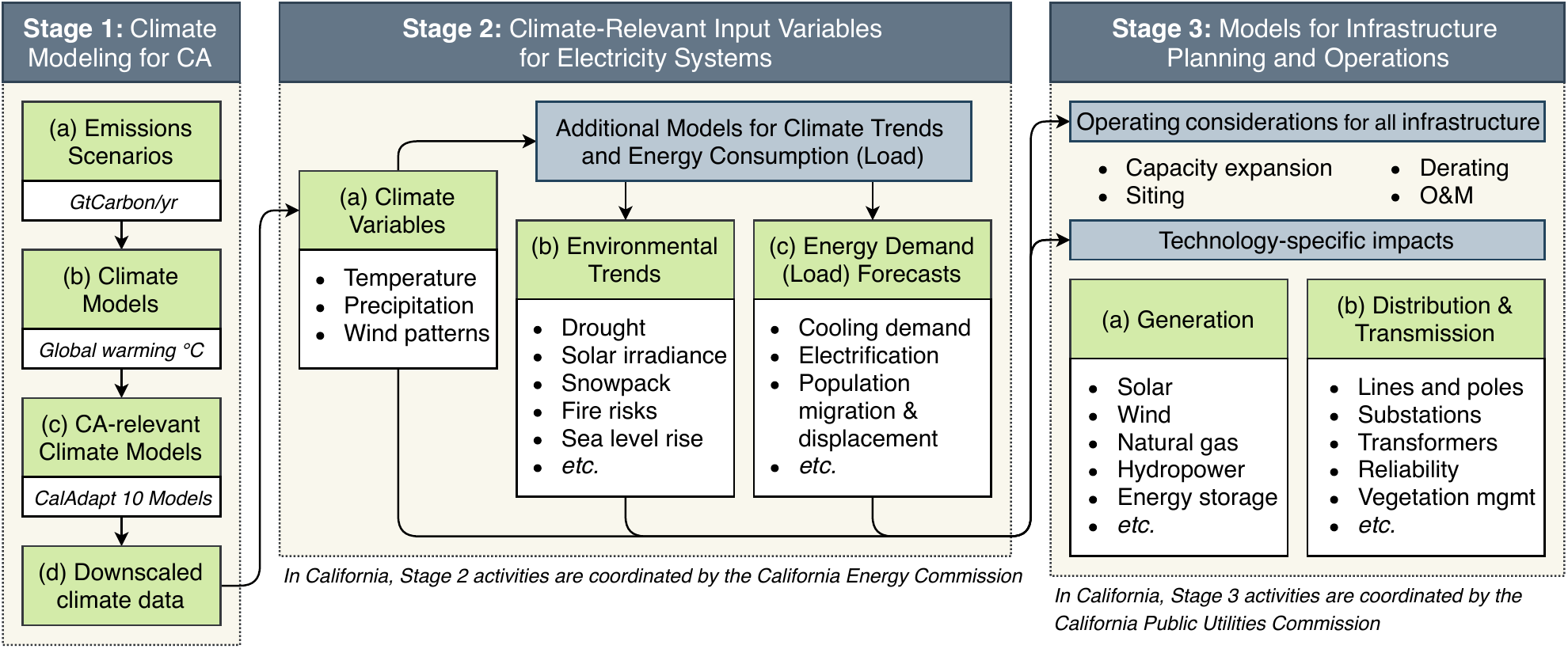}
\caption{Flow of information between climate models and grid infrastructure planning. Each box represents a distinct modeling effort. This diagram is not intended to be comprehensive, and additional data streams may exist.}
\label{fig:flowchart}
\end{figure}

In other sectors, agencies have established guidelines and best-practices for incorporating climate information into decision-making processes (see, for example, \cite{ipcc_planning, LHC2014, opr2018}). 
These efforts identify and  quantify climate vulnerabilities and outline possible mitigation strategies.
Though simple decision analysis models may be suitable in some planning contexts, the severe consequences associated with failure to detect unmitigated risks in electric power systems suggests that comprehensive analysis of climate impacts is warranted \cite{shi2019}.
This analysis will require operationalizing climate projections, quantifying impacts on environmental and operating characteristics relevant to power systems, and evaluating near- and long-term implications of grid operations and planning decisions.
Figure \ref{fig:flowchart} illustrates these information flows.
Collaboration between scientists and practitioners will be necessary \cite{briley2015, Gerlak2018}, and state or federal planning authorities can play a critical role in coordinating how information are acted upon by different types of decision-makers. 
%Interpreting the results of these models to inform l
Long-term investment decisions will ultimately need to assess the costs associated with mitigating climate risks and the ramifications of allocating limited resources to mitigate certain risks but not others. The societal implications of possible risk scenarios (i.e., wildfires, widespread blackouts, rising energy costs), motivate the need to incorporate climate information into assessments of infrastructure vulnerabilities that are (or will be) present.
Effective policy can prevent duplication of efforts, educate practitioners about the nature and the limitations of existing climate data, and define best practices.

 % flowchart

The current work synthesizes what is known now about the interactions between grid infrastructure and climate and provides recommendations for moving forward. 
We focus on a case study of California: a state that has already done a great deal of research to develop data and guidelines to begin incorporating climate considerations into decision-making processes.
%We summarize the work that has been done, and comment on the work that remains.
Our contributions are threefold: 
First, we provide background on climate models, grid planning models, and decision-making processes that inform if, how, and where investments are made. 
Next, we review the factors that must be considered for future planning and discuss how the models listed in Figure \ref{fig:flowchart} would need to evolve to quantify grid/climate interactions in a detailed and comprehensive way.
Finally, we offer recommendations for decision-makers in California's electric sector to begin to act upon climate projections.
With these recommendations, we 
%shed light on the work that remains to be done, and 
identify specific actions that researchers, practitioners, and policymakers can take to ensure that our understanding of climate risks to grid infrastructure continues to advance.

We provide background on electricity infrastructure planning and climate adaptation efforts in California in Section \ref{sec:currentplanning}. Then, following the flow of information in Figure \ref{fig:flowchart}, we provide background on climate modeling (Stage 1 activities) in Section \ref{sec:climatemodeling}. The climate-relevant input variables for electricity systems listed in Stage 2 are introduced and discussed in Section \ref{sec:trends}. Considerations for translating those climate inputs into infrastructure planning models are discussed in Section \ref{sec:infrastructure} (covering generation, distribution, and transmission). We provide overarching recommendations in Section \ref{sec:recommendations} and conclude in Section \ref{sec:conclusions}.

% CALIFORNIA BACKGROUND
\section{The California planning context: Electricity systems and climate adaptation}
\label{sec:currentplanning}

\subsection{Electricity system planning}

In California, electric utilities share primary responsibility for energy- and electricity-related planning and oversight with two state agencies and the state's independent grid operator (\ref{sec:appCAplanning}).
The California Energy Commission (CEC) generates hourly demand forecasts for its Integrated Energy Policy Report (IEPR).
The California Public Utilities Commission (CPUC) and regulated utilities use these forecasts to identify investments needed to continue to provide reliable, safe, and cost-effective electricity service.
Demand forecasts also inform transmission planning decisions overseen by the California Independent Systems Operator (CAISO).

The IEPR forecast includes scenarios related to weather, energy efficiency, and load growth futures \cite{iepr2017}.
Planning decisions that must be robust to extreme weather events are informed by a 1-in-10 weather year generated from statistical analysis of historic data \cite{Burillo2017}.

Two of the IEPR scenarios include climate adjustments for ``low'' and ``high'' temperature rise scenarios.
Further documentation is necessary to understand how these adjustments account for changes in consumption for each end use \cite{Orford2019}.
However, the documentation suggests that adjustments are based on the temperature-sensitivity of existing load and do not consider more profound changes in energy consumption (see Section \ref{sec:energyuse}). 
Because the IEPR load forecasts are used throughout the state to inform planning decisions, incorporating a rigorous assessment of climate impacts here could help decision-makers account for climate impacts in a consistent and coordinated manner \cite{Knapstein2019}.

Aside from load impacts, modeling processes that inform infrastructure planning do not account for climate trends. 
\ref{sec:appCAplanning} contains a thorough discussion of existing infrastructure planning processes in California and the data flows between them. 
In many cases, modeling assumptions and architecture may need to be revised to comprehensively factor in existing climate-grid interactions (Section \ref{sec:infrastructure}).

% These forecasts are used as inputs to the models the California Public Utilities Commission (CPUC) and regulated utilities use to identify investments needed to continue to provide reliable, safe, and cost-effective electricity service.
% Demand forecasts also inform transmission planning decisions overseen by the California Independent Systems Operator (CAISO). 
% In an effort to ensure that the system is operable during all but the most extreme events, current models rely on historical data. Data are sampled and weighted to capture both typical and extreme operating conditions. 
% For example, a 1-in-10 historical weather year may inform certain planning decisions \cite{Burillo2017}. 
% Climate-adjusted load forecasts report trends in energy use and peak load under `low' and `high' temperature increase scenarios. 

%The CPUC oversees and approves these investments for recovery through the rates that residential, commercial, and industrial customers pay for electricity.
% Evolving trends in extreme operating conditions due to climate change largely do not factor into planning decisions.

\subsection{Climate change adaptation}
\label{sec:currentplanning:climate}

The State of California initiated climate research and planning efforts in the late 1980's; these efforts are detailed in \cite{Orford2019} and \cite{LHC2014}. Here, we summarize recent and ongoing efforts specifically related to adaptation and infrastructure. 
These include research into potential climate impacts, efforts to operationalize climate data, and legislative and regulatory directives for planning agencies. 

California solicits adaptation research through the state's climate change assessments, a joint effort by the Governor's Office of Planning and Research, the California Natural Resources Agency, and the CEC. Starting in 2006, the state has completed four rounds of assessments.
These examine physical vulnerabilities, adaptation options, and research needs \cite{CA2019ClimateAssessmentweb}. 

Executive Order S-13-2008 initiated strategic planning processes for sea level rise and climate adaptation \cite{ExecOrdS132008}, resulting in guidance documents in both areas. 
The sea level rise guidance document (updated in 2018) provides a synthesis of state-of-the-art science on sea level rise, and outlines best-practices for coastal adaptation \cite{SLRguidance2018}. 
Notably, the document takes a step-wise approach to setting risk tolerances, and acknowledges that these may differ across categories of decision-making. For example, an ``extreme" scenario is included for consideration for ``high-stakes, long-term decisions" \cite[p.4,25]{SLRguidance2018}.
The climate adaptation guidance document, issued in 2009, underscores the importance of a data clearinghouse---namely CalAdapt, then under development---to ``synthesize existing California climate change scenarios and climate impact research'' \cite{cnra2009}.\footnote{Additional resources hosted by the CA Governor's Office of Planning and Research include datasets and planning guidelines for climate adaptation by local planning agencies (e.g., see: \cite{resilientCA}).}

Legislative actions in 2015 and 2016 built on these efforts: Senate Bills 379 and 246 and Assembly Bills 1482 and 2800 established mechanisms to coordinate adaptation efforts, required state agencies to consider climate change while planning for state infrastructure, and directed stakeholders to use CalAdapt data when assessing local climate vulnerabilities \cite[p.117]{CEC2016IEPRupdate}. 
One initiative formed out of this legislation was the Climate-Safe Infrastructure Working Group, which recommended adaptive planning, whereby decision-makers move forward with currently-available information while taking note of information gaps \cite{CSIWG2018}. 
The working group emphasized the need to confront changes in both average and extreme weather conditions \cite{CSIWG2018}, and to use probabilistic methods to deal with uncertainty in risk \cite{CSIWGDeane2018appendix7}. 
Our own recommendations build on these principles (Section \ref{sec:recommendations}).

In April 2018, the CPUC opened a rulemaking focusing on climate change adaptation \cite{CPUC2018OIRClimateAdaptation}. Designed to integrate climate awareness into grid infrastructure planning and regulatory decision-making statewide, the proceeding asked stakeholders to suggest approaches, data sources, and tools to ``address climate adaptation in a consistent manner" \cite{Randolph2018}. 
Here, we contribute to this effort by examining climate information, modeling needs, and decision processes specific to adapting electric grid infrastructure to maintain safe and reliable service under evolving climate conditions.

% CLIMATE MODELING BACKGROUND
\section{Background on climate modeling}
\label{sec:climatemodeling}

Climate models use assumptions about worldwide emissions to predict possible short- and long-term trends in weather variables.
%Global emissions scenarios describe the amount of carbon injected into the atmosphere. 
%They are the primary inputs to physics-based models that characterize climate dynamics.
%As output, these models predict possible short- and long-term trends in weather variables under a given emissions scenario.
We do not know precisely how emissions or climate dynamics will play out over the next century. 
State-of-the-art climate projections use a range of modeling assumptions designed to capture different possible climate futures.
%Statistical or numerical analysis of these outputs (referred to as downscaling) is often necessary to generate data granular enough to inform planning purposes.
The different stages of the climate modeling process are described in detail below.

\subsection{Global emissions scenarios}
\label{sec:climatemodeling:emissions}
Emissions scenarios describe the amount of carbon and other greenhouse gases injected into the atmosphere. 
The Intergovernmental Panel on Climate Change (IPCC) releases data for a range of possible emissions scenarios, or Representative Concentration Pathways (RCPs). 
These are numbered by the increase in radiative forcing (a metric of global warming, reported in W/m$^2$), and include RCP2.6, 4.5, 6.5 and 8.5 W/m$^2$.

Climate data localized to  California are readily available for RCP8.5 and RCP4.5.
RCP8.5 describes a ``business as usual'' trajectory where emissions increase at current rates for the remainder of the century.
In RCP4.5, emissions increase over the next 50 years, then decrease to below 1990 levels by 2100.

These two scenarios were selected during a 2015 analysis of climate information for state water resources planning \cite{cctag}, when available data for other emissions scenarios were not as comprehensive.
Climate projections are continuously being updated to reflect new emissions pathways and advances in our understanding of climate dynamics. 
Decisions about which data to use must be periodically revisited as climate science advances.

\subsection{Climate models}
Climate projections come from global circulation models (GCMs) that characterize the physical dynamics driving circulation and heat transfer between the Earth's atmosphere, land, oceans, and ice caps.
As these interactions are not perfectly understood, teams of scientists have developed a library of GCMs that incorporate modeling assumptions designed to capture different possible dynamics, reported in \cite{ipcc_climatemodels}.
Over 50 climate projections for different GCMs and emissions scenarios are consolidated in the Coupled Model Intercomparison Project (CMIP) \cite{cmip5}.

A 2015 study mined the most recent models (CMIP5) to identify suitable projections to inform water resource planning in California \cite{cctag}.
Fifteen models were found to accurately characterize regional weather systems. Of these, ten accurately predicted precipitation metrics particularly relevant to water resource planning \cite{cctag}, and were chosen for wide dissemination through CalAdapt \cite{caladapt}.
The relevance of these models to support applications beyond water resources (e.g., grid infrastructure planning) remains to be examined.

The CEC selected a smaller subset of four models to support the Fourth Climate Change Assessment \cite{cec2018b}. These models cover a similar range of temperature and precipitation outcomes as the 10 models.
In 2017, the CEC recommended that these four models (with the RCP4.5 and RCP8.5 emissions scenarios) should be used for energy sector planning \cite[p.137]{CEC2016IEPRupdate}. 
In 2019, the CPUC directed electric utilities to use the 10 GCMs available within CalAdapt (with RCP8.5) for decisions related to planning, investment, and operations \cite{Randolph2019ClimAdaptDec}.

\subsection{Downscaled climate models}
\label{sec:climatemodeling:downscaling}

The spatial resolution of GCM outputs is very coarse (250 to 600 km), making raw climate projections ill-suited for use in granular planning decisions. 
%applications sensitive to regional weather variations, or where characterizing extreme values (rather than averages) is critical.
A mathematical process called ``downscaling'' is used to infer variability in weather conditions within large GCM grid cells to estimate changes at finer geographic scales.
Temporal downscaling may also infer variability on finer time-scales.
Downscaling is necessary to generate data suitable to inform most applications, particularly those where climate impacts are sensitive to regional weather variation, and where planning decisions must be robust to extreme (rather than average) conditions.

There are two common approaches for downscaling GCM data.
\textit{Dynamic downscaling methods} use parametric models to approximate physical dynamics that give rise to regional variation.
\textit{Statistical downscaling methods} mine historical weather data to quantify regional variability, and generate projections exhibiting similar statistical properties.
Dynamic downscaling is generally considered to be more accurate and better-suited to characterizing extremes, but it is very computationally intensive and may be biased by boundary conditions or imperfect understanding of physical dynamics.
Statistical downscaling methods require less computational power, but assume that statistical properties of local weather phenomena will remain constant (or stationary) in spite of climate trends.
This assumption is known to be false \cite{kunkel2013}.
These differences and practical considerations for choosing a particular downscaling approach are discussed broadly in \cite{barsugli2013}, and in the context of water resource planning in \cite{cctag}.
The CEC commissioned research to develop state-of-the-art statistical downscaling methods for California, known as Localized Constructed Analogs (LOCA) \cite{pierce2014, Pierce2018}.
LOCA downscaled versions of 10 climate models are available through CalAdapt.
% In addition to downscaled projections of weather variables (e.g., temperature, precipitation), CalAdapt includes results from a number of studies projecting climatic trends related to weather (e.g., drought, wildfire risk, stream flow, snowpack) and sea level rise.
% This repository is intended to provide a standard library of climate data for researchers and planning agencies across California to use \cite{cnra2009}.
% The California Governor's Office of Planning and Research provides additional resources offering more targeted support to decision-makers on how to incorporate climate data into planning decisions.
% These resources include a guidebook to help local agencies identify and plan for potential climate impacts \cite{opr2018}, and the Adaptation Clearinghouse, which points users to resources including climate data and case studies.

% INPUT VARIABLES
\section{Climate-relevant input variables for electricity systems}
\label{sec:trends}

A number of climate trends will impact grid infrastructure.
Chronic impacts, like sea level rise, more rapid equipment aging, and increasing electricity demand, will stress existing infrastructure over time. 
Acute impacts, like wildfires and severe weather, will lead to much more sudden consequences. 
Anticipating these effects can improve strategies for mitigating risks and responding to emergencies.

Using climate projections as inputs to infrastructure planning models (Section \ref{sec:infrastructure}) can ensure that planning decisions are robust to the chronic and acute vulnerabilities.
Here, we summarize relevant climate trends. 
We consider three types of inputs (referring to stages 2a, 2b, and 2c, respectively, in Figure \ref{fig:flowchart}): \textit{climate variables}, which are direct outputs from climate models, \textit{environmental trends}, which are derived from climate variables, and \textit{energy demand}, which is heavily influenced by weather and climate. 
Table \ref{tab:caladapt} summarizes data available through CalAdapt about climate variables and trends.

%Planning for new infrastructure will also differ from accounting for existing assets. New infrastructure can be designed, sited, and operated to account for future expected climate impacts. Adapting existing assets will require more complicated decisions about hardening and early retirement and/or replacement. Planning scenarios must consider opportunities related to new infrastructure options as well as account for existing infrastructure. 

\subsection{Climate variables}
\label{sec:trends:vars}

These variables can be obtained directly as outputs from climate models or at higher resolution from downscaled LOCA models.
The following paragraphs describe electric power system impacts.

\paragraph{Temperature}
\label{sec:trends:temperature}

Climate models largely agree that temperatures will rise and heat waves will become more frequent and more intense \cite{Ward2013}. 
Rising temperatures will impact load growth, generator efficiency, equipment ratings, and degradation rates (among other factors) \cite{EPRI2009,Bartos2015}.
Extreme heat events may have cascading effects \cite{Clark2019}. 
% \begin{itemize}\setlength\itemsep{-0.1em}
% \item Increased electricity demand for cooling and air conditioning (AC) in the summer, including in locations with little AC penetration to-date, which may lead to significant spatial shifts in load;
% \item Reduced generation efficiency of thermal plants due to operating cycle thermodynamics and increased water cooling requirements (which may be further exacerbated by increases in surface water temperature); 
% \item Reduced generation efficiency of solar photovoltaics (PV) due to internal voltage reductions in PV cells \cite{Dubey2013};
% \item Reductions in the operating capacity of existing distribution and transmission infrastructure, including sagging transmission lines on hot days; and,
% \item More rapid aging and more frequent failures for equipment, including transformers.
% \end{itemize}
%We elaborate on these impacts in the following sections.

\paragraph{Precipitation}
\label{sec:trends:precipitation}

The direction and magnitude of projected trends in precipitation vary across climate models. 
Models agree, however, that seasonal and spatial variability will increase, affecting power generation, heat dissipation, and maintenance needs. 
Greater variability makes characterizing extreme events critically important, as operational and planning decisions are often informed by extreme precipitation events \cite{Voisin2016}. 
%Less precipitation may limit access to water for cooling in plant operations, and necessitate more frequent cleaning of PV panels and other equipment. 
%But extreme precipitation events are also expected to increase and may bring other problems, such as flooding and associated risks to infrastructure \cite{Ward2013}. 
%Voisin \textit{et al.} proposed metrics to describe the potential impacts of water availability on power systems planning with particular focus on thermoelectric and hydropower generators \cite{Voisin2018}.

\paragraph{Wind speed}
\label{sec:trends:wind}

Expected temperature increases and pressure changes in atmospheric currents as a result of climate change will have a direct impact on wind patterns. However, research is needed to determine how these changes will impact regional wind patterns relevant to grid planning decisions \cite{Craig2018}. Climate change is expected to affect ``the intensity and duration of sustained winds" \cite{EPRI2009} and to increase peak wind intensity \cite{Ward2013}. These trends will impact the operation and performance of both wind turbines and the infrastructure that must withstand winds (i.e., power lines). Wind speeds are also a crucial input for calculating the potential heat impacts on equipment, as wind can provide cooling to offset high temperatures \cite{Bartos2016}.

\subsection{Environmental trends}
\label{sec:trends:envt}

These trends cannot be obtained directly from climate models, but instead follow from changes in the intensity, geography, and seasonality of climate variables discussed above. Projections are generated by analyzing climate model outputs.

\paragraph{Drought} 
\label{sec:trends:drought}

Increasing temperatures and changing precipitation patterns may lead to more frequent and severe droughts. 
Drought will increase electricity demand associated with water pumping for drinking, irrigation, and other uses \cite{EPRI2009}, and could introduce additional loads for desalination.
Severe drought events may carry additional ramifications for power sector operations in California \cite{Zohrabian2018}.

\paragraph{Solar irradiance}
\label{sec:trends:irradiance}

Changes in temperature and precipitation will impact atmospheric conditions that drive variables like humidity and cloud cover (occurrence, type, timing, and optical thickness) \cite{Craig2018}. Changes in these conditions will impact surface solar radiation, thereby affecting solar generation, net load from rooftop solar PV, and the apparent temperature on the ground.

\paragraph{Snowpack}
\label{sec:trends:snowpack}

Warmer winters at high altitudes will lead to more precipitation falling as rain rather than snow and an earlier melting time for snowpack. These factors may lead to reduced water availability during the summer months due to the changing timing of runoff. Many reservoirs in California are dual-purpose: they were built to accommodate water from slow-melting snow into the summer months, and include extra storage capacity for flood control. Increased precipitation will lead to earlier snowmelt, which will increasingly coincide with the flood season. An increase in water released to protect against floods in the spring will reduce water availability through the summer \cite{CADWR2015}. Hydro resources will be further impacted as snowpack disappears from lower elevations \cite{EPRI2009}. Models predict that the Sierra snowpack may decrease by 48-65 percent by 2100 from its 1961-1990 average \cite{CADWR2015, Rhoades2018}.

\paragraph{Fire risk}
\label{sec:trends:fire}

Increasingly warm, dry, and windy conditions may exacerbate existing wildfire risks \cite{PGE2018}. Reduced snowpack and earlier snowmelt may also lengthen the wildfire season. The impact of these climate variables on fire risk may be further exacerbated by modern fire suppression practices \cite{Wahl2019}. Wildfire risks include damage to grid infrastructure, and the need to pre-emptively de-energize lines to prevent ignition \cite{EPRI2009,Dale2018}.

\paragraph{Sea level rise}
\label{sec:trends:sealevel}

Sea level rise impacts on infrastructure can include coastal flooding, coastal erosion, exacerbated land subsidence, saltwater intrusion, and pipeline corrosion \cite{DeAlmeida2016}. In the context of electricity infrastructure, sea level rise will predominantly impact siting decisions and expected damages to coastal infrastructure and facilities, including generating plants and substations \cite{EPRI2009,Bruzgul2018}. Sea level rise impacts may also extend beyond coastal areas as rivers swell and low-lying areas resist drainage after high tides.

\begin{table}[p]
    \centering
\begin{tabular}{>{\raggedright\arraybackslash}p{1.5in} >{\raggedright\arraybackslash}p{3in} >{\raggedright\arraybackslash}p{1.48in}} 
    \hline \noalign{\medskip}
\multicolumn{1}{c}{\textbf{CalAdapt Data Stream}} & \multicolumn{1}{c}{\textbf{Description}} & \multicolumn{1}{c}{\textbf{Planning Relevance}} \\  \noalign{\medskip} \hline \noalign{\medskip}
Raw LOCA Downscaled Climate Data &  Daily projections of relative humidity, surface solar radiation, and wind speed available through the CalAdapt data server; data are more challenging to interface with than the data streams listed below & Solar capacity; Grid hardening; Planning for Extremes \\ \noalign{\smallskip}
        
Annual Averages\footnotemark[1] & Annual minimum and maximum temperatures; Total annual precipitation & Peak capacity; Derating; Planning for extremes \\ \noalign{\medskip}
        
Cooling and Heating Degree Days\footnotemark[1] & Degree-day estimates derived from difference between daily minimum and maximum temperature and user-defined heating/cooling setpoint temperatures & Load forecasting; Capacity expansion \\ \noalign{\medskip}
        
Extended Drought\footnotemark[2] \cite{Pierce2018} & Weather/hydrologic projections for two extreme drought scenarios (early \& late century) & Hydro capacity; Water availability for power plant cooling; Planning for extremes \\ \noalign{\medskip}
        
Extreme Heat Days and Warm Nights\footnotemark[1]$^,$\footnotemark[3] & Frequency and intensity of hot days/nights for various ``extreme'' event thresholds & Peak capacity; Derating; Reliability; Planning for Exremes; Load forecasting; Siting \\ \noalign{\medskip}
        
Extreme Precipitation\footnotemark[1]$^,$\footnotemark[3] & Frequency and intensity of precipitation events for various ``extreme'' event thresholds & Hydro capacity; Distribution reliability; Storm hardening \\ \noalign{\medskip}
        
Hourly Projections of Sea Level\footnotemark[2] \cite{kopp2014,DeConto2016} & Projects sea levels associated with diurnal/seasonal tidal patterns and arctic ice melt & Siting \\ \noalign{\medskip}
        
Sea Level Rise (CalFloD-3D)\footnotemark[2] \cite{caladapt_sealevelrise} & Projects sea level inundation during 100-year storm events at high spatial resolution for the Bay Area, San Joaquin River Delta, and California Coast & Grid hardening; Siting; Planning for extremes \\ \noalign{\medskip}
        
Snowpack\footnotemark[2] \cite{Livneh2015} & Monthly snow water equivalent & Hydro capacity \\ \noalign{\medskip}
        
Streamflow\footnotemark[2] \cite{caladapt_streamflow} & Monthly and annual streamflow projections for 11 streamflow gauging stations throughout the state of California & Hydro; Siting \\ \noalign{\medskip}
        
Variable Infiltration Capacity (VIC) Variables\footnotemark[1] & Provides a wide range of hydrologic variables with daily resolution & Hydro capacity \\ \noalign{\medskip}
        
Wildfire\footnotemark[2]$^,$\footnotemark[4] \cite{caladapt_wildfires} & Provides 5- and 10-year averages of acres burned under different population growth scenarios & Siting; Grid hardening; Capacity expansion; Planning for Extremes \\\noalign{\medskip}
        
    \hline \noalign{\smallskip}
    
\multicolumn{3}{p{5.98in}}{\footnotemark[1]\footnotesize{Data are derived directly from statistical processing of LOCA downscaled climate model outputs.}} \\
\multicolumn{3}{p{5.98in}}{\footnotemark[2]\footnotesize{Data are generated from VIC variables and/or other data sources. Relevant documentation are cited where appropriate.}} \\
\multicolumn{3}{p{5.98in}}{\footnotemark[3]\footnotesize{Though data are reported for each 6km grid cell, the statistical methods used to estimate extreme events are meant to broadly describe changes across the state. Additional analysis may be needed to provide actionable information to decision-makers on a local scale.}} \\
\multicolumn{3}{p{5.98in}}{\footnotemark[4]\footnotesize{Data are available for the subset of 4 climate projections used in the Fourth Climate Change Assessment, not for all 10 climate models.}} \\
% \multicolumn{3}{p{5.98in}}{\footnotemark[4]\footnotesize{References cited report historic streamflow estimates; no specific documentation is provided about how streamflow projections are generated.}} \\ \noalign{\smallskip} \hline
    \end{tabular}
\caption{Summary of data streams available through CalAdapt and their relevance to grid infrastructure planning. All data streams report climate projections from 2006-2100; many also report historical data for 1950-2006. Data are reported for LOCA downscaled models at 1/16th of a degree (about 6km$^2$) spatial resolution for the entire state of California. References to the studies that produced each data stream are listed, except where data are directly computed from LOCA downscaled climate model outputs (see footnote 1 above). We refer readers to CalAdapt \cite{caladapt} for additional detail about data contents and units of measure.}
    \label{tab:caladapt}
\end{table}

% ENERGY CONSUMPTION AND LOAD
\subsection{Energy demand}
\label{sec:energyuse}

% Warmer temperatures will increase demand for cooling and ventilation loads.
% Temperature-sensitive loads will become more intensive on a per household basis, and more extensive, as space conditioning becomes a necessity in regions that previously experienced milder temperatures \cite{auffhammer2018}.

% Drought, fire risk, and other trends discussed in Sections \ref{sec:trends:vars} and \ref{sec:trends:envt} could also change energy consumption patterns.
% Meanwhile, meeting California's aggressive emissions targets will require broad electrification of end uses such as heating and transportation \cite{williams2012}.
% Trends contributing to an increase in load may also be offset by advancements in energy efficiency.

Climate will impact energy consumption and the generation, distribution, and transmission resources needed to reliably serve these evolving demands.

\paragraph{More intensive load}
\label{sec:energyuse:cooling_demand}
Rising temperatures will increase the electricity drawn by existing uses. For example, electricity demand for cooling, refrigeration (particularly in warehouses), and other loads that maintain thermal comfort (e.g., ventilation and fans) will grow.
Higher peak temperatures and more frequent extreme heat days will induce higher and more frequent peak load events \cite{Beard2010}.
These trends will increase overall energy consumption, thereby increasing base load requirements, and may affect seasonal load patterns.
The specific regional impacts will depend on the characteristics of local building stock \cite{Dirks2015}.
Extreme heat will also raise the stakes for power outages during peak load events, as space conditioning becomes a necessity for vulnerable populations \cite{Pincetl2016}.

\paragraph{More extensive load}
\label{sec:energyuse:cooling_penetration}
Climate change will also increase electricity demand from new uses and in new locations. 
For example, warming will lead to more extensive cooling demand in historically moderate climate zones (e.g., San Francisco).
More extensive space conditioning will increase system peaks and could stimulate more consistent demand for cooling throughout the year.
This trend could impact decisions related to generator siting and capacity expansion in transmission and distribution networks.
The IEPR forecast currently relies on appliance saturation data last collected in 2009 \cite{rass2009}, and methodological revisions to the forecast may be warranted to ensure that changes in appliance saturation and trends in ownership are included.

\paragraph{Electrification of new end uses}
\label{sec:energyuse:decarbonization}
Achieving California's aggressive emissions targets will require electrifying end uses---such as manufacturing, heating, and transportation---that are currently served by fossil fuels.
This will increase base load and alter diurnal and seasonal load shapes. For example, electrification of heating loads could prompt wintertime peak load events.
%Though climate forecasts predict that winters will become more moderate, extreme weather events may also become more frequent and more severe.
%While extreme weather poses a threat to grid reliability, 
Heightened reliance on electricity for heating and transportation could make the implications of wintertime power outages increasingly severe. Meanwhile, extreme weather events that threaten grid reliability may become more common.

\paragraph{Population displacement}
\label{sec:energyuse:population_displacement}
The above trends in load are expected given current population trends, but climate change may induce additional shifts in population and therefore electricity demand.
A recent report estimates that sea level rise alone will displace over 250 thousand people nationally and 30 thousand in California by 2100 \cite{Scientists2018}.
Displacement due to drought, natural hazards, and conflict will add to these numbers, and could occur on much shorter timescales.
Migration to urban areas could increase electricity demand in existing load pockets.
Migration to less-populated regions may warrant expansion of transmission and distribution infrastructure, and may also increase the wildland-urban interface, thereby putting more people at risk of power shutoffs and further displacement due to wildfires \cite{strikeforce2019}. 
Population growth due to economic factors unrelated to climate change may also exacerbate these trends.

% INFRASTRUCTURE IMPACTS
\section{Electricity infrastructure planning models and impacts from a changing climate}
\label{sec:infrastructure}

\begin{figure}[b]
    \centering
    \includegraphics[width=0.9\linewidth]{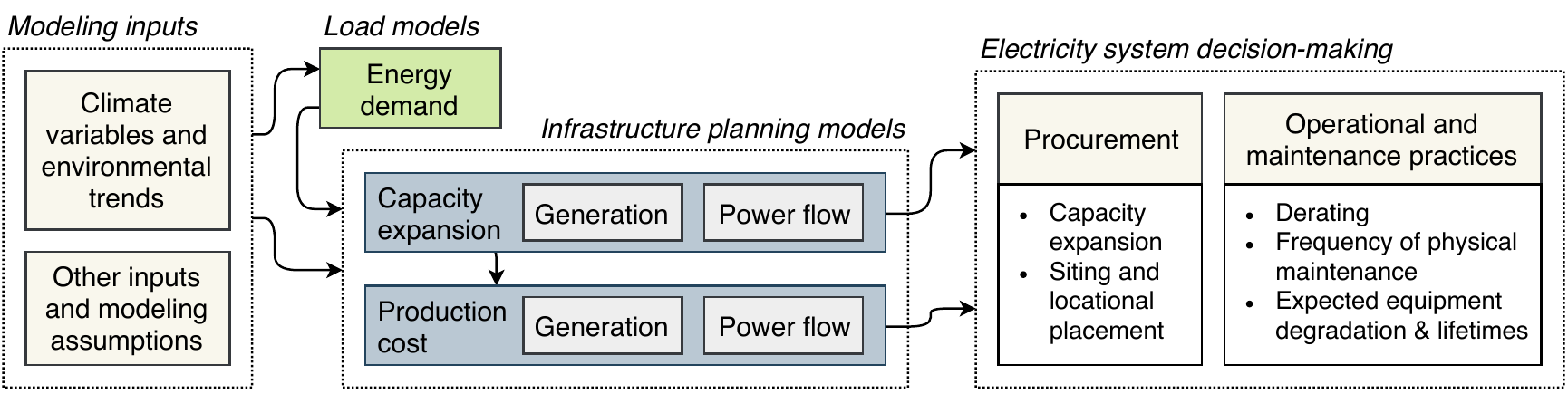}
     \caption{Modeling stages for electricity infrastructure planning.}
    \label{fig:IPMs}
\end{figure}

Five primary types of models inform electricity system planning decisions. These include: 

\begin{itemize}\setlength\itemsep{-0.1em}
\item \textit{Generation models} simulate the physical operation of specific power generation technologies. 
\item \textit{Power flow models} describe how electricity moves through wires between generators and consumers. 
\item \textit{Load models} project how trends in population, energy use intensity, and other factors will impact the magnitude, shape, and geographic distribution of energy use.
\item \textit{Capacity expansion models} optimize generation and transmission procurement decisions based on load forecasts, capital costs, and operating assumptions.
\item \textit{Production cost models} simulate how grid assets can meet demand, reliability, and emissions requirements at least cost given operating constraints.

\end{itemize}
These models use different assumptions and inputs. 
Figure \ref{fig:IPMs} outlines the flow of information between them.
 %IPMs
Climate trends introduced in Section \ref{sec:trends} will affect grid infrastructure performance.
To anticipate climate vulnerabilities, infrastructure planning models will need to account for these trends.

Here, we synthesize what is known about how climate trends will alter grid infrastructure performance. 
We focus on component- and system-level impacts on operational power systems; for a discussion of resilience and recovery from outages, we refer readers to \cite{Arghandeh2016, Wang2016}.
Where possible, we reference previous work to provide a sense of the magnitudes of the impacts described. 
We focus on climate impacts most relevant to grid infrastructure in California, and refer readers to \cite{Craig2018, CA2019ClimateAssessmentweb} for a more comprehensive review.
Table \ref{tab:climateimpactsummary} summarizes key takeaways from our assessment of the potential impacts.

% IMPLICATIONS FOR ELECTRICITY GENERATION
\subsection{Electricity generation}
\label{sec:generators}

These considerations apply to both centralized and distributed energy resources (DERs). 
While statewide resource planning currently focuses on transmission-level bulk generation, increasing DER deployment in California is prompting conversations about how smaller-scale resources could meet localized generation needs and relieve pressure on existing infrastructure \cite{Arghandeh2014, CPUC2017DERActionPlan}.

\subsubsection{System-wide impacts}

Climate trends will impact the design and operational performance of generation in regional power systems.
We comment on changes in grid modeling capabilities needed to characterize these trends. 

\paragraph{Capacity expansion}
\label{sec:generators:capacity}

Decisions to add generation are informed by capacity expansion models that issue recommendations about the mix of generation resources appropriate to supply load. 
Climate impacts on load shapes (Section \ref{sec:energyuse}), and generator performance (Section \ref{sec:generators:tech}) could alter planning recommendations.

Existing models for evaluating resource adequacy in local generation fleets rely on IEPR load forecasts modified to a wider range of weather scenarios \cite{CaliforniaPublicUtilitiesCommission2019} (\ref{sec:currentplanning:load}).
Scenarios are currently based on historical weather data, though similar adjustments could informed by climate forecasts \cite{Rogers2019,Young2019}.
Climate trends may lead us to revisit planning reserve margins. A recent study indicated that existing reserve margins may become inadequate by the end of the century due to temperature rise \cite{Burillo2017}. 

\paragraph{Siting}
\label{sec:generators:siting}

Sea level rise may require that we retrofit or relocate low-elevation facilities, and make it harder to build new generation in coastal areas.
Water scarcity may also influence design decisions for water-cooled generation facilities. 

Siting decisions may also improve resilience to contingencies caused by more frequent severe weather events (e.g., wildfires, storms, heat waves).
% Heat waves in particular cause demand to increase, while simultaneously reducing the transmission capacity of grid infrastructure.
% These coinciding effects increase the risk of outages; the presence of local generation resources can mitigate that risk. 
For example, local generation resources may enable grid operators to de-energize power lines to mitigate fire risk without interrupting critical loads and services.
Planning models traditionally have not examined interactions between ambient conditions and generator siting decisions. 

% Models characterizing these benefits should account for projected trends in load, extreme weather, and wildfire risk, as well as the reliability implications of these trends.
% Although existing models examine generation resources from many different perspectives (e.g., reliability, cost, emissions), to our knowledge the modeling infrastructure that could capture compounding benefits (e.g., between wildfire mitigation and resource adequacy) does not currently exist.

\paragraph{Generator derating}
\label{sec:generators:derating}

Derating refers to the practice of operating equipment below its maximum rated capability to avoid internal damage to equipment or external damage to the environment.
Derating may be necessary to prevent material degradation of power generators (and grid components) as ambient temperatures become more extreme. 
These changes may warrant new operating and maintenance practices, additional generation capacity, or energy efficiency and load management programs. 

\paragraph{Compounding impacts}
\label{sec:generators:compounding}

Changes in equipment performance warrant new grid modeling capabilities to capture system impacts that may be greater than indicated by component-level analyses.
For example, derating---which impacts generation, distribution (Section \ref{sec:distribution:derating}), and transmission (Section \ref{sec:transmission}) equipment---is necessary during extreme heat waves, when the system is also more likely to be under operational stress (due to peak load events). Under the same conditions, power outages pose a public health risk (due to extreme outdoor air temperatures), and the generation capacity of solar and thermal generators decreases (Section \ref{sec:generators:tech:solar}).
Current planning models do not fully consider the temperature-dependence of operating constraints.
This assumption should be revisited to ensure that operating limits are not inadvertently exceeded to reduce the risk of correlated failures during extreme heat events.

\subsubsection{Implications for generation technologies}
\label{sec:generators:tech}
Specific technologies will bear climate impacts due to their physical characteristics.
%Here, we discuss climate impacts on specific generation technologies.
% Technology-specific considerations impact how climate change variables will affect the operation and power production of energy generating facilities in California.

\paragraph{Solar photovoltaics}
\label{sec:generators:tech:solar}
Solar generation improves with increasing irradiance but declines with temperature (see \ref{sec:appGen:PV}) \cite{Dubey2013}. 
To evaluate generation capacity in a warming climate, system models will need to account for PV derating.
%Existing grid planning models consider changes in generation due to irradiance, but not temperature.
A recent study estimated that PV capacity in the Western U.S. could decrease by approximately 0.7-1.7 percent by mid-century due to higher temperatures \cite{Bartos2015}.

Less precipitation may also compel more frequent manual cleaning of PV panels to preserve operational efficiency.

\paragraph{Wind}
\label{sec:generators:tech:wind}
Wind speed patterns influence decisions related to siting, design, and operations of wind farms.
Changes in average and extreme wind speeds will alter the performance of existing wind farms (see \ref{sec:appGen:wind}), and may alter optimal siting for new installations \cite{Craig2018}.
Researchers have identified a need to improve wind speed projections \cite{Craig2018}. Studies to date suggest that wind generation potential in California could decrease \cite{Pryor2012}.

% Turbines are sized based on the predicted wind characteristics at a particular site. 
% Power generation is sensitive to the actual conditions a turbine faces: it scales in proportion to wind speed cubed, but ceases completely when conditions exceed a set cut-out wind speed to prevent equipment damage (\ref{sec:appGen:wind}). 

% Climatic trends could alter the statistical properties of wind fields throughout the state, including short-term (inter- and intra-day) variability and long-term (annual) averages \cite{Craig2018}.
% These changes could reduce wind generation from existing wind farms, or may increase potential power production at sites where wind speeds are not currently favorable. 
% Increases in the frequency of extreme wind speed events could also reduce power production by increasing the amount of time wind turbines need to be shut off because conditions exceed the cut-off wind speed \cite{Beard2010}.
% Notably, while existing wind farms may increasingly face conditions unsuitable for the particular choice of turbine sited in that location, technology development could help mitigate some of these issues for newer wind fields \cite{Craig2018}.

% Recent studies reflect the persistence of large uncertainties in climate impacts on wind patterns, and more research is needed \cite{Craig2018}. 
% Though researchers have identified a need to improve wind speed projections \cite{Craig2018}, studies suggest that the West may experience a decline in wind generation potential \cite{Pryor2012}.

\paragraph{Thermal power generation}
\label{sec:generators:tech:ng}
Power production from thermal generators relies on a temperature differential between high-pressure steam (heated by combustion) and ambient water (or air) (\ref{sec:appGen:naturalgas}) \cite{Peer2016}.
These generators---which include natural gas, concentrating solar, and some biomass plants---are more efficient when this temperature differential is high.
Rising ambient temperatures will reduce plant  efficiency and generating capacity \cite{Bartos2015,Voisin2018,Miara2017}. 
Plants at risk of violating thermal discharge limits may need to be curtailed \cite{Cook2015, McCall2016}.
Heat waves will exacerbate these issues at times when generation resources are needed to meet increased cooling demand \cite{Ke2016}. 

Other factors may also impact performance.
Drought will limit water availability, potentially causing water-cooled plants to become inoperable for part (or all) of the year \cite{Cook2015}.
Changes in load may increase peak load relative to total demand, reducing the capacity factor of thermal generators (which typically serve peak load). 
Operating at less than full capacity will reduce the efficiency of thermal plants \cite{Craig2018}.

The capacity of thermoelectric plants in the Western Interconnection is estimated to decrease 1.6-3.0 percent on average by mid-century, not including losses attributable to drought \cite{Bartos2015}. 

\paragraph{Hydropower}
\label{sec:generators:tech:hydro}
Changes in snowpack, as well as the seasonality and amount of precipitation will impact reservoir levels and water availability \cite{Beard2010} (Section \ref{sec:trends:snowpack}). 
Drought will reduce hydropower generation \cite{Voisin2018}. 
Intense precipitation may lead reservoir operators to prioritize flood control, making hydro resources less readily available to support grid needs \cite{EPRI2009, Tarroja2016, Zhou2018}. 
Thus although its fast-ramping capabilities make hydro well-suited to provide frequency response, other generation technologies (e.g., DERs, thermal generators) may need to provide these services in the future. 
Less snow and earlier melting times may reduce annual hydropower generation in California up to approximately 3 percent \cite{Tarroja2016}.

\paragraph{Energy storage}
\label{sec:generators:tech:storage}
Energy storage---including electrochemical storage, pumped hydropower, and other emerging \linebreak technologies---can balance intermittencies in renewables generation. Local storage resources may also provide islanding capabilities should transmission or distribution equipment become inoperable, for example due to physical damage or high wildfire risk. 
Ambient temperature may alter battery degradation and performance; grid impacts, however, have not yet been studied.

% Batteries can support islanding of regional power networks that become inoperable due to damage, overloading, or high wildfire risk.
% Thus batteries can provide valuable grid services related to both climate change mitigation and adaptation.

% Storage can play a critical role in climate adaptation. Stored energy can be dispatched much like conventional generators during peak load events, and (with proper islanding capabilities) can supply loads when grid service is lost due to extreme weather or deenergization events. Energy storage will also play a critical role in climate mitigation by managing renewable generation intermittency and reducing curtailment. Distributed energy storage may be able to mitigate the need for grid upgrades to accommodate high penetrations of rooftop solar. A full understanding of the benefits batteries can provide, however, will require models examining capacity expansion, distribution system planning, real-time energy markets, and reliability and risk due to severe weather and wildfires.

% DISTRIBUTION INFRASTRUCTURE
\subsection{Distribution infrastructure}
\label{sec:distribution}

The distribution system carries electric power from substations to end-use customers through a diverse array of equipment and lines. 
Transformers convert electricity from high- to low-voltage, and feed it to customers through overhead lines (often held up by wooden poles) or underground cables. 
Along the way, voltage regulators, capacitor banks, circuit breakers, and other equipment enhance power quality, resilience and safety. 
Here, we discuss climate impacts on distribution systems planning and operations.

\subsubsection{Derating of distribution grid components}
\label{sec:distribution:derating}
Distribution grid components are designed to operate under a specific range of loading conditions, determined by properties of the constituent materials.
These limits involve heat dissipation, and relate to internal cooling mechanisms and ambient temperatures. 
Different ratings may apply at different operational timescales (e.g., continuous operation versus temporary load spikes or instability).

Components are sized to meet peak loading conditions (with some safety margin).
Optimal design may involve operating components at or near their rated limits during peak load, as excessive safety margins may lead to undue costs \cite{ren2008}.
Historic data and load forecasts inform design decisions. 
Warming trends will increase peak loads and decrease heat dissipation---thus restricting safe loading limits, particularly for transformers and overhead lines \cite{Ward2013,Panteli2015, Burillo2018}.
Failure to derate components (and operate the system in adherence with those ratings) as ambient temperatures increase could lead to more rapid degradation, increasing failure rates, and general reductions in equipment lifetimes \cite{Beard2010}.
A recent study estimated that distribution components in Los Angeles could experience a 2-20\% loss of rated capacity by 2060 due to heat waves, increasing the risk of overloading components in congested areas \cite{Burillo2018}.

Climate variables besides temperature can also impact power ratings. For example, long periods of dry weather can reduce the thermal conductivity of the ground, requiring further derating of underground cables \cite{Ward2013}. Changes in moisture may also reduce the efficiency of earthing at substations, requiring additional safety precautions \cite{Panteli2015}.

% Derating power flow limits to account for increasing temperatures can therefore help support system resiliency and equipment lifetimes: ``Thermal limits on components are more restrictive on hot days. If components are not derated to allow for this, they may fail more frequently, age faster, and require more maintenance and earlier replacement" \cite{Beard2010}.

%"The direct effect of an increase in average temperature on transmission or distribution is to limit or reduce the maximum power rating of equipment and to increase the energy losses. The impact is similar to the effect of a small increase in power flow. Hence, the mitigation of higher temperature in the future requires just a small extension to the measures used routinely to accommodate increasing power flows due to growth in consumer demand and changing generation patterns." \cite{Ward2013}

% Practically, planning for thermal equipment derating on a particular circuit is similar to accounting for load growth and increasing power demand \cite{Ward2013}: in either case, the circuit may need upgrades and/or adjustments to the timing and magnitude of power flows, which may be aided by distributed energy resources and storage. In this case, existing equipment may also require updated operating protocols. 

\subsubsection{Siting}
\label{sec:distribution:siting}

Derating needs will depend on localized temperature conditions.
Population growth in hot areas of the state means that derating could become a concern for a larger share of grid components.

Distribution equipment may also be impacted by flooding and sea level rise. 
Flooding during extreme precipitation events, for example, will impact equipment in low-lying areas---in particular, switchgear, control cubicles, and transformers at ground level in substations \cite{Ward2013, Panteli2015}. 
Similarly, sea level rise will impact coastal substations and other equipment. 
A recent study found potential impacts to four substations in San Diego Gas \& Electric's service territory, as well as ``thousands of electric substations, transformers, power lines, and other equipment [that] are potentially exposed to damage under scenarios of sea level rise" \cite{Bruzgul2018}.

\subsubsection{System design and connectivity}
\label{sec:distribution:capacity}
System upgrades expanding the capacity of distribution systems may be warranted to accommodate new load and compensate for equipment derating.
Changes in the connectivity of distribution systems (e.g., islanding, load shedding, and enhanced sectionalization) may also support grid operations during capacity shortfalls or when wildfire risk is high \cite{Xu2015}.
Islanding capabilities and local generation resources can ensure that critical loads maintain service continuity during outages \cite{Arghandeh2014}.
In regions of the state where climate risks make it cost-prohibitive to build safe, robust and reliable grid infrastructure, the obligation to serve may be better met by permanently islanded microgrids \cite{Li2017, Bajwa2019}.

\subsubsection{Operations \& maintenance}
\label{sec:distribution:oandm}

Climate variables and trends pose various challenges for distribution system maintenance practices and reliability. Specific examples highlight the need for more frequent inspections and careful maintenance to support system performance:
\begin{itemize}\setlength\itemsep{-0.1em}
\item Rising temperatures directly contribute to equipment aging. For example, faster chemical degradation of insulating materials directly increases the failure rates of conductors and transformers \cite{zhang2007a}.
\item Increased loading and power flows (Section \ref{sec:energyuse}) results in additional stress ``as operational conditions approach thermal and mechanical ratings of power system elements," leading to greater ``overall wear and tear" and ``increased vulnerability to faults and/or breakdowns" \cite{Beard2010}.
%"These conditions may contribute to deterioration of dielectric materials, operating mechanisms, supporting structures, and cooling/insulating liquids." \cite{Beard2010}
\item Heavy rain can damage overhead lines, and soak equipment such as insulators and switchgear increasing the risk of short-circuit and arcing faults. These issues can be mitigated by newer equipment designs and careful maintenance \cite{Ward2013}.
%"Very heavy rain occasionally causes flashover faults (short-circuits) across insulators, but changes to the design of insulators can reduce the risk of this (EPRI 2007). Heavy rain can cause water ingress into high voltage insulators and switchgear, leading to internal flashovers and catastrophic failures, but careful maintenance can minimise the risk of such faults." \cite{Ward2013}
\item Precipitation poses longer-term risks to distribution systems. For example, moisture leads to internal decay of wood poles, reducing structural integrity \cite{Shafieezadeh2014}. This in turn puts conductors at greater risk.
\item Changing wind patterns and extreme wind gusts could threaten overhead lines, network towers, and other overhead structures. High winds impose shear force on poles and towers, and increase the likelihood that vegetation or other debris will cause damage \cite{Ward2013}, or lead to faults \cite{Beard2010,Panteli2015}. Frequent inspections and hardening tower and pole designs to withstand stronger winds could mitigate these impacts \cite{Ward2013}.
%"If more extreme wind gusts occur, they would cause tower and conductor damage and more faults due to galloping and trees falling." \cite{Beard2010}
%"High winds during storms and hurricanes can lead to faults and damage to overhead transmission and distribution lines, either by debris being blown against the lines or even a tower collapse in extremely high winds." \cite{Panteli2015} 
%"The combination of rain with strong wings or lightning can however be significant threat to overhead lines." \cite{Panteli2015}
%"Direct wind damage to the network towers and poles and substation structures, is minimised by designing them to withstand the maximum expected wind loadings (e.g., BSI 2001), and by frequent inspection of their integrity." \cite{Ward2013}
\item Wildfires (as well as the intense winds that often accompany them) can cause physical damage to distribution infrastructure and increase maintenance needs \cite{Dale2018}. Smoke is also a concern, as a high concentration of ions makes smoke more conductive than air, increasing the risk of arcing on overhead lines \cite{Ward2013}.
\end{itemize}

\subsubsection{Vegetation management}
\label{sec:distribution:vegetation}
A safe distance must be maintained between grid infrastructure and vegetation.
Contact is a common cause of power outages and can also cause infrastructure damage, arcing, or tree ignition that---when ambient conditions are appropriate---can spark wildfires \cite{SDGE2018}.
%Climate-aware tree trimming policies will need to account for evolving fire risk.
High temperatures coupled with heavy loading during heat waves can cause overhead lines to sag. Clearance can be maintained by modifying vegetation management practices, or in some cases by derating lines.
Extreme wind speed events increase the risk of contact from falling and swaying tree limbs.

Climate-aware vegetation management policies will need to consider temperature, wind speed, seasonal patterns that influence tree growth (e.g., length of growing season and ecosystem health), and eventually also changes in tree and shrub species that surround grid infrastructure.
Longer growing seasons may warrant more frequent tree trimming \cite{Ward2013}, while ecosystem damage due to aridification or invasive species (such as bark beetles) may warrant the removal of trees that are in poor health \cite{SDGE2018,PGE2018}.

\begin{table}[p]
    \centering
    \begin{tabulary}{1\linewidth}{l L L L} \hline \noalign{\medskip}
\textbf{Climate Impacts} & \multicolumn{1}{c}{\textbf{Generation}} & \multicolumn{1}{c}{\textbf{Distribution}} & \multicolumn{1}{c}{\textbf{Transmission}} \\ \noalign{\medskip} \hline \noalign{\medskip}
Temperature & Solar and natural gas: rising temperatures reduce efficiency of power production & Derating and increased line losses, more rapid equipment aging & Derating and increased losses, increased congestion \\
\noalign{\bigskip}
Precipitation & Hydro: reduced energy generation
capacity, less flexible dispatch & Water inundation risks for equipment, faster aging for wooden poles, changes to vegetation management & Changes to vegetation management \\
\noalign{\bigskip}
Wind patterns & Wind: reduced power production from existing farms, possible creation of new sites & Equipment damage, changes to vegetation management, fire ignition and spread & Equipment damage, changes to vegetation management \\ \noalign{\medskip} \hline \noalign{\medskip}
Drought & Natural gas: less water for cooling; Hydro: less water for power production & Reduced soil thermal conductivity requiring derating of underground cables & \\
\noalign{\bigskip}
Solar irradiance & Solar: stronger irradiance increases power production & & \\
\noalign{\bigskip}
Snowpack & Hydro: reduced energy generation capacity, less flexible dispatch & & \\
\noalign{\bigskip}
Fire risks & Energy storage: supports ``non-wires alternatives'' for mitigating fire risk & Equipment damage and increased risk for arcing faults, changes to vegetation management & Equipment damage and resiliency impacts due to line outages \\
\noalign{\bigskip}
Sea level rise & Siting: relocation of existing assets, design challenges for new generation, corrosion & Water inundation and corrosion risks for equipment & Water inundation and corrosion risks for equipment \\ 
\noalign{\medskip} \hline \noalign{\medskip}
Load & Possible need for capacity expansion to serve additional peak and base load & System stress and increased maintenance requirements; Possible need for capacity expansion & Possible increased congestion \\ \noalign{\medskip} \hline
    \end{tabulary}
    \caption{Summary of key potential climate impacts on electric infrastructure in California.}
    \label{tab:climateimpactsummary}
\end{table} % climateimpactsummary

% TRANSMISSION INFRASTRUCTURE
\subsection{Transmission infrastructure}
\label{sec:transmission}

The transmission system carries power from large generators over long distances to regional load centers. 
In California, CAISO conducts an annual planning process to address evolving system needs (\ref{sec:currentplanning:TPP}). 
Many of the inputs rely on historical weather data. %, thereby propagating problematic inputs through planning for the state's bulk energy infrastructure.
%In this work, we focus primarily on the distribution system because it is regulated locally by the CPUC and managed by distribution utilities in the state. 
While California's transmission system is operated by CAISO, it is regulated by the Federal Energy Regulatory Commission (FERC). 
Though federal regulation is beyond the scope of this paper, we briefly comment on climate interactions with the transmission system.

Similar to the distribution system, transmission lines are assigned a power rating for maximum electricity transfer. 
Transmission lines that approach this maximum power rating are said to be `congested' and act as bottlenecks in moving electric power from one area to another. 
Higher ambient temperatures reduce heat dissipation from transmission lines, thus increasing energy losses and reducing line transfer limits  \cite{Bartos2016,Panteli2015}. 
Heat-related capacity reductions will likely coincide with peak loading, further stressing the system and prompting concerns about supply adequacy \cite{Bartos2016}.
As in distribution systems, failure to account for temperature-dependence of operating constraints could lead to correlated component failures.
In transmission systems, such correlated failures may have cascading effects that result in widespread blackouts \cite{clarfeld2019}.

A recent study estimated that projected temperature increases during the month of August could reduce transmission line transfer limits in California by 7-8 percent by 2100 \cite{Sathaye2013}. 
Technological solutions exist: for example, heat-resistant cables provide new options for preserving and improving line capacity, but it remains to be seen whether re-conductoring lines is best solution \cite{Bartos2016}. 
Challenges associated with transmission system expansion may eventually lead to greater reliance on local generation.
%"High temperatures and heat waves limit the transfer capability of transmission lines, and increase the energy losses and the line sagging." \cite{Panteli2015}

Increasing fire risk in California also threatens transmission facilities \cite{Panteli2015}. A recent study found that distribution infrastructure incurred more damage than transmission infrastructure from 2000-2016 wildfires in California \cite{Dale2018}. However, damage to transmission facilities and/or preventative de-energization due to nearby wildfires or risky fire-prone weather \cite{Dale2018} may impact a large number of customers. Moreover, changing climate conditions may make transmission (and distribution) equipment more likely to trigger fires.
One option for mitigating risk is to invest in infrastructure upgrades that reduce the probability of component failures. Another alternative is to modify the topology of the network to remove transmission lines from regions of the state where wildfire risk is excessively high.

% RECOMMENDATIONS
\section{Recommendations}
\label{sec:recommendations}

%may be useful sources here for policy and risk mgmt: \cite{Gerlak2018} and \cite{Nierop2014}

Effective climate adaptation measures will require decisions that are informed by known interactions between climate science and power systems.
Both are highly technical areas, and many of the interactions between them are not yet well-understood.
Moreover, climate risks are not necessarily represented in historic data. 
Projections of variables characterizing these risks are uncertain at best, and, for certain variables, do not yet exist in a form that decision-makers can readily use.
Here, we describe opportunities for progress, in light of these existing challenges.
% Climate change adaptation for electricity systems will require making both long-term and real-time decisions informed by complicated interactions between two extremely complex systems: climate science, and grid planning and operations.
% In many cases these interactions are not yet well understood. Moreover, the information needed to study them further are uncertain at best, and may not yet exist in a form decision-makers can readily use.
% This undertaking is formidable, yet there are clear opportunities to make progress.

\begin{figure}[tbh]
    \centering
    \includegraphics[width=0.9\linewidth]{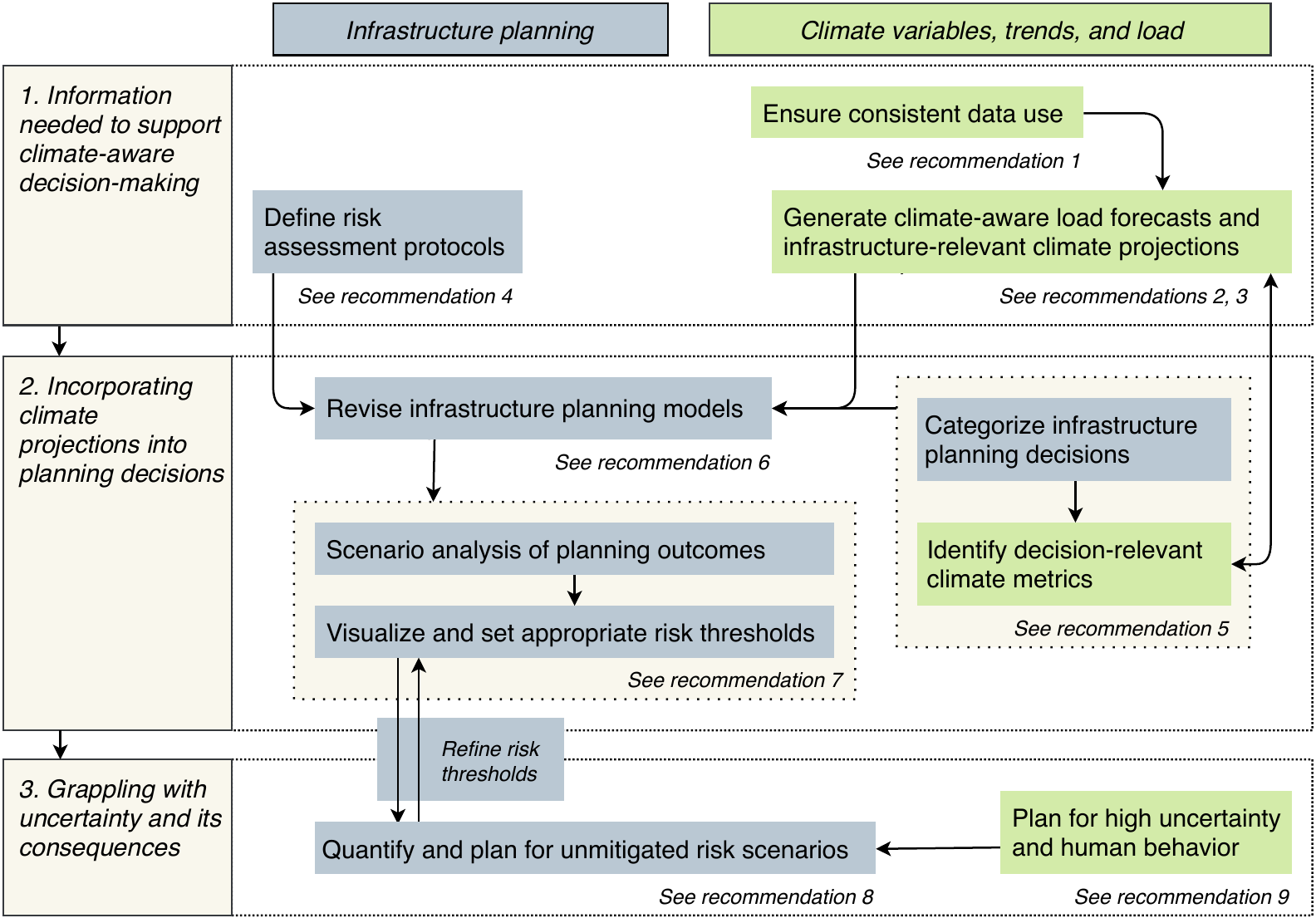}
    \caption{A proposed process for incorporating climate information into infrastructure decision-making.}
    \label{fig:decisionprocess}
\end{figure}

We share nine recommendations for applying best-practices in climate adaptation to grid infrastructure planning processes in California.
By following these recommendations, decision-makers may begin to account for grid-climate interactions in a \textit{comprehensive} manner.
By \textit{comprehensive}, we mean that planning decisions consider grid-climate interactions, compounding effects, and associated impacts on diverse stakeholder groups--including utilities, ratepayers, and disaster relief agencies.
We build on best-practices in adaptive planning, including: defining well-documented risk tolerances \cite{SLRguidance2018}, accommodating advancements in climate science \cite{CSIWG2018}, and revisiting adaptation strategies as new information comes to light \cite{Haasnoot2013}.
Together, our recommendations provide a roadmap for decision-makers to better understand and act upon climate risks, as illustrated in Figure \ref{fig:decisionprocess}. 
Each recommendation offers near-term actions that can be pursued in parallel today. 
First, we describe the information needed to support climate-aware decision-making (Recommendations 1, 2, 3, and 4). Then, we describe how these can inform infrastructure planning models and decisions (Recommendations 5, 6, and 7). Finally, we discuss the need for decision-makers to internalize uncertainties inherent to planning for climate change and its consequences (Recommendations 8 and 9).

\subsection{Information needed to support climate-aware decision-making}
\label{sec:recommendations:info}

To begin, adaptation efforts must be informed by data reflecting the current state of climate science. Where relevant, climate data may need to be tailored to the specific case of grid infrastructure planning.

\subsubsection*{\textbf{\textsc{Recommendation 1:} Regulators should specify which climate data stakeholders are to use to inform specific planning decisions.}}

\paragraph{\textbf{\textsc{Barrier}}} 
Stakeholders that use and disseminate climate data must grapple with challenges such as:
\begin{itemize}\setlength\itemsep{-0.1em}
\item Climate science is evolving, and user-facing data must be kept up to date. 
\item Climate model assumptions may make certain data ill-suited for certain applications; data limitations are far from intuitive and require detailed understanding of both the climate models and the application of interest.
\item Substituting historic data with climate projections as inputs to existing planning models may not be appropriate, and may not account for all of the interactions that are present.
\end{itemize}
%A key barrier to modeling climate impacts is that comprehensive data sets and modeling assumptions are not readily available or widely agreed upon. 
In California, CalAdapt currently provides data that can readily support climate-aware planning (Table \ref{tab:caladapt}). However, additional work remains to be done to evaluate whether these data: (1) remain consistent with state-of-the-art climate science, and (2) provide requisite information to inform grid infrastructure planning decisions (data requirements are also discussed in Recommendations 2-4). Moreover, stakeholders will require clear direction on how to apply this data to inform their decisions (discussed in Recommendations 5-7). 

\paragraph{\textbf{\textsc{Solution}}}
%Clear guidelines must be provided about which data sets to use and how to use them. Without clear direction, stakeholders may take different approaches to data selection and post-processing.
In California, CPUC should specify which climate data will inform planning decisions to ensure that different stakeholders use consistent information, and that these information remain up-to-date.
CPUC should form a technical advisory group to determine how relevant data will inform specific decisions. 
\textbf{\textsc{In the near-term,}} the CPUC should instruct stakeholders to start with data that currently exists via CalAdapt. 
% The CPUC should explicitly state which specific CalAdapt data sets and versions thereof are relevant and should be accepted as standard in its existing proceedings. 
\textbf{\textsc{Ultimately,}} CEC should work with stakeholders to assess whether CalAdapt should offer additional data streams to support grid planning needs.
For example, filters for selecting relevant GCMs (discussed in \cite{cctag}) and downscaling methods (discussed in \cite{barsugli2013}) may need revisiting to provide requisite information for grid planning applications.
CEC should ensure that the data to support these applications is readily available (see Recommendations 2 and 3), and release updates as new data become available due to advancements in climate science. 
% CPUC should provide guidelines detailing which data to use and how to use them to inform planning decisions.

\subsubsection*{\textbf{\textsc{Recommendation 2:} Model climate impacts on load, and generate a library of load forecasts that span all relevant climate scenarios.}}

% \subsubsection*{\textbf{\textsc{Recommendation 2:} Develop a portfolio of load forecasts that more thoroughly account for climate trends across available climate scenarios.}}

\paragraph{\textbf{\textsc{Barrier}}} 
As discussed in Sections \ref{sec:currentplanning} and \ref{sec:energyuse}, as well as \ref{sec:currentplanning:load}, IEPR load forecasts examine a limited range of climate scenarios, and do not consider climate impacts on end use saturation and consumption behaviors. 

\paragraph{\textbf{\textsc{Solution}}} 
Since electricity infrastructure planning relies heavily on IEPR forecasts, revisiting these forecasts will have impacts that propagate through the planning process.
\textbf{\textsc{In the near term,}} CEC can support other grid planning entities by providing load forecasts for all climate projections available in CalAdapt. 
Methods and assumptions for generating end-use forecasts should be clearly documented to ensure that merits and limitations of the forecasts are transparent, and to help downstream analysts determine if or where post-processing is warranted.
\textbf{\textsc{Ultimately,}} the CEC will need to revise load forecasting methods and assumptions to more thoroughly account for nuanced interactions between energy use and weather (beyond temperature), and to account for mitigation and adaptation measures taken by both policy-makers and end-users (e.g., electrification of new end uses and higher saturation of air conditioners).
Engaging with stakeholders who rely on IEPR forecasts to inform their work may reduce the need for ad-hoc post-processing by grid planning entities.

\subsubsection*{\textbf{\textsc{Recommendation 3:} Identify climate data relevant to grid infrastructure planning, and conduct research necessary to generate data that do not already exist.}}

\paragraph{\textbf{\textsc{Barrier}}} 
Though data exist to support certain types of planning decisions (e.g., water resource use \cite{cctag} and natural gas pipeline siting \cite{caladapt_sealevelrise}), additional data may be needed to support new applications like grid infrastructure planning.
Data reporting temperature, precipitation, sea level rise are relevant to multiple applications, but other vital datasets are lacking.
For example, information about the frequency, severity, and seasonality of high-wind speed events are critical to understanding wind power production and to informing grid hardening measures.

\paragraph{\textbf{\textsc{Solution}}} 
Imperfect information need not be a barrier to making climate-aware decisions; planning efforts can begin to move forward with the data currently offered in CalAdapt.
\textbf{\textsc{In the near term:}} Where possible, infrastructure planners should use climate projections.
CEC should engage with stakeholders to understand the limitations of existing data, and to identify additional data requirements (see Recommendation 5).
\textbf{\textsc{Ultimately:}} Research is needed to generate a comprehensive library of data streams for grid infrastructure planning. 
Research needs include determining whether statistical or numerical downscaling methods are more appropriate, and to project regional changes in frequency and severity of high-wind speed events.
% For example, while statistical downscaling may be adequate for certain planning decisions, numerical downscaling methods may be essential for others. 
% Research is also warranted to predict regional changes in frequency and severity of high-wind speed events from climate projections.
CEC should support these efforts, while working with CPUC and other stakeholders familiar with data requirements unique to supporting grid planning applications.

\subsubsection*{\textbf{\textsc{Recommendation 4:} Develop quantitative methods to project risk exposure of infrastructure assets and assess mitigation options.}}

\paragraph{\textbf{\textsc{Barrier}}} 
Robust risk assessment protocols must evaluate the probability and implications of scenarios that may never have occurred previously.
Risk reduction measures must balance diverse trade-offs---such as service reliability, equity, and loss of life. 
Further study is required to assess the societal implications of different adaptation strategies.
% RAMP filings demonstrate that both utilities and CPUC are in the process of grappling with these challenges (\ref{sec:currentplanning:risk}).
In California, utilities already use probabilistic methods to assess risks and inform upgrade decisions \cite{sed2018, sed2015}.
However, incorporating climate projections to forecast climate-related risks to grid assets has not yet been discussed \cite{sed2018}.

\paragraph{\textbf{\textsc{Solution}}} 
CPUC must ensure that utilities use models that can examine evolving risks due to climate change.
Methods for evaluating these risks must be transparent and subject to public scrutiny.
Existing regulatory proceedings in California (including S-MAP and RAMP, see \ref{sec:currentplanning:risk}) provide an avenue for assessing risks, but currently consider only certain types of risks.
\textbf{\textsc{In the near term,}} CPUC should engage with stakeholders to identify metrics appropriate for characterizing climate risks.
Safety-focused risks covered in existing regulatory proceedings (e.g., wires down, fire ignitions), should be considered alongside broader risks (e.g., reliability and equity).
New metrics for characterizing climate risks should also be included (e.g., inundation of grid assets due to sea level rise).
CPUC should oversee the process of developing quantitative methods for anticipating new and evolving risks, for example by mining historic data to quantify climate-sensitivities of risk metrics.
\textbf{\textsc{Ultimately,}} risk assessment models should quantify how proposed investments will impact risk and performance over the lifetime of new and existing assets.
These models should be used to optimize candidate investments, and identify strategies that minimize overall risk exposure (where possible) or meet specified risk thresholds (see Recommendation 7).
The magnitude and severity of damages due to possible risk scenarios (e.g., wildfires) should also be explicitly reported.
Transparent reporting will allow utilities, regulators, and communities to weigh the implications of mitigating certain risks and not others, and can ensure that stakeholders understand and agree to shoulder the implications of unmitigated risks (see Recommendation 8).

\subsection{Incorporating climate projections into planning decisions}
\label{sec:recommendations:planning}

With a quantitative basis for examining climate trends and evolving risks, we can begin to develop infrastructure planning processes that internalize complex interactions between climate and infrastructure to inform planning decisions.
Here, we propose a potential approach. 
Recommendations 5 and 6 focus on scoping and implementation, respectively, while Recommendation 7 synthesizes results into action.

\subsubsection*{\textbf{\textsc{Recommendation 5:} Perform a comprehensive assessment of potential grid-climate interactions.}}

\paragraph{\textbf{\textsc{Barrier}}} 
Electricity infrastructure planning decisions are informed by many different physical, operational, and societal considerations. The decisions are complex, and no individual stakeholder group will be familiar with all possible implications of different adaptation strategies.

\paragraph{\textbf{\textsc{Solution}}} 
A comprehensive mapping of how climate variables, trends, and load may affect specific infrastructure decisions is a necessary prerequisite to incorporating these factors into infrastructure planning models. 
CPUC should engage a wide variety of subject-matter experts in this effort---for example in climate science, hydrology, power generation and distribution equipment, grid design, operations and repairs, energy use consumption, and power systems modeling---through a technical advisory or working group. 
Here, we propose an approach for mapping out these interactions: 
\begin{enumerate}
    \item List specific infrastructure planning decisions that must be made.
    \item List climate impacts that could influence each decision. Table \ref{tab:climateimpactsummary} provides a starting point based on the interactions discussed in this paper; engaging with a broader audience of stakeholders and experts could shed light on additional interactions.
    \item Define how the magnitude of each climate-grid interaction will be quantified (e.g., instantaneous impacts, cumulative exposure, etc.).
    \item Define metrics for decision-making specifying: (1) relevant climate variable(s), (2) appropriate statistics (e.g., low, extreme, average, etc.) and duration, (3) geography, and (4) decision timescale  \cite{Jagannathan2019drcm}. Examples could include water inundation (to inform decisions to relocate certain assets) or extreme wind speed projections (to inform pole reinforcement).
% For example, a decision-relevant climate metric for assessing maintenance strategies to combat the risk of equipment water inundation may be the projected maximum precipitation per day in the city of Sacramento in 2040. Similarly, decision-relevant climate metrics for identifying the best way to serve a 10\% increase in residential customers connected to a particular substation may include maximum and minimum hourly load forecasts for the next 20 years, estimates of high hourly temperatures in residential areas (to account for line derating), and extreme wind speed projections (for pole reinforcement).% The development of these metrics should occur with the aid of an advisory group and/or stakeholder process.
\end{enumerate}

\noindent\textbf{\textsc{In the near term,}} the CPUC should form a technical advisory group to enumerate climate impacts on planning decisions (as detailed above). Results should be circulated for public comment, and revised as appropriate. \textbf{\textsc{Ultimately,}} this document should be refined to comprehensively map out climate impacts on specific planning decisions. Mitigation options should be weighed based on input from stakeholders ranging from ratepayers to repair personnel. %Stakeholders can then access this climate data for input into downstream planning models via CalAdapt.

\subsubsection*{\textbf{\textsc{Recommendation 6:} Refine infrastructure planning models to incorporate climate impacts.}}

\paragraph{\textbf{\textsc{Barrier}}} 
The models we currently use to inform infrastructure planning decisions do not comprehensively account for grid-climate interactions. 
Existing models contain baked-in assumptions about operations, maintenance, and performance that may not hold as climate and loading conditions change.
Planning models must be revised to account for new climate realities. 

\paragraph{\textbf{\textsc{Solution}}}
Well-informed and climate-aware recommendations will require infrastructure planning models to thoroughly incorporate sensitivities between infrastructure performance and climate trends, such as those listed in Table \ref{tab:climateimpactsummary}.
%Infrastructure planning models need to be adapted to analyze climate and load scenarios, and to incorporate dynamic assumptions about climate variables and trends.  
\textbf{\textsc{In the near term,}} CPUC should lead efforts to ensure that planning models can run multiple climate scenarios to examine how different futures could influence today's planning decisions. 
Modeling infrastructure may need to be adapted to ingest new input data as CMIP releases updated climate projections to reflect scientific advances. 
\textbf{\textsc{Ultimately,}} planning models should be revised such that embedded assumptions are dynamic and account for changes in performance due to emerging climate trends. New models that account for compounding interactions may need to be developed. Risk calculations (see Recommendation 4) should be embedded in infrastructure planning models to ensure that decisions focus not only on maintaining grid operations, but also on mitigating risks.

\subsubsection*{\textbf{\textsc{Recommendation 7:} Determine the range of planning outcomes across different climate projections. Set appropriate risk thresholds.}}

\paragraph{\textbf{\textsc{Barrier}}} Infrastructure planning models must be adapted to examine numerous climate scenarios, climate impacts of varying magnitudes, and uncertainty inherent in making decisions informed by climate projections.
Furthermore, climate impacts may compound; for example, capacity expansion models must examine temperature-sensitivities of both load growth and equipment ratings.

\begin{figure}[b!]
\centering
\includegraphics[width=0.9\linewidth]{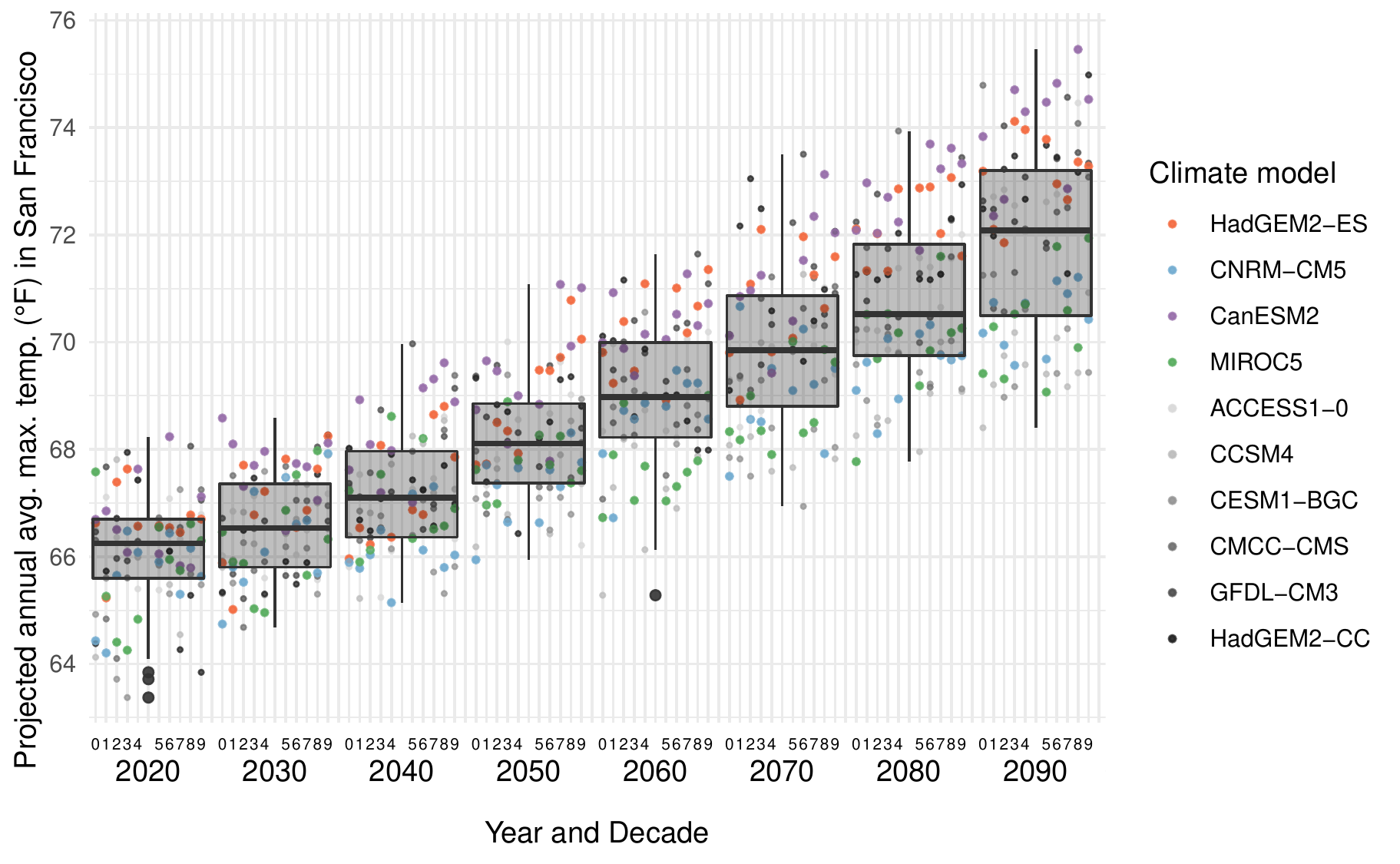}
\caption{An example box-and-whisker plot. Data points indicate the projected annual average maximum temperature within the San Francisco city limits from each of the 10 California-relevant climate models available via CalAdapt under scenario RCP 8.5 \cite{CalAdapt2019SFmaxtemp}. Boxplots are drawn by decade (i.e., years 2040 through 2049 are assigned to 2040) and centered over the underlying data points.
%These data indicate temperature projections from each climate model for each available year. 
Temperature projections indicate the range of environmental conditions that infrastructure could face during each decade. Outliers are defined as points further than 1.5 times the interquartile range from the 25th or 75th percentile \cite{Wickham2019}. 
For more granular planning decisions, box-and-whisker plots could be drawn for (e.g.) a single year from 10 predictions, one from each model. 
The first four climate models listed have been recommended as priority models for research purposes and broadly represent `hot/dry,' `cool/wet,' `average,' and `most distinct' conditions, respectively \cite{cec2018b}. The other models are listed alphabetically.}
\label{fig:gcmbxplt}
\end{figure}

\paragraph{\textbf{\textsc{Solution}}} Decision-makers must consider the range of planning options given different climate outcomes.
Examining the differences (or lack thereof) will provide insight into the implications of planning to different risk thresholds.
Based on the results, CPUC should engage with stakeholders to define appropriate risk tolerances for specific planning decisions.
Explicitly setting a risk tolerance will ensure that planning decisions are robust to uncertainty in climate projections, and can provide transparency needed to balance reduced costs against the possibility of incurring additional risks (see Recommendation 8). 
One way to approach this challenge is as follows:
\begin{enumerate}
    \item Run infrastructure planning models for each climate scenario specified in Recommendation 1; identify recommended investments for each scenario. (A sensitivity analysis is advisable, as sensitivities to individual climate variables may vary.)
    \item Visualize the range of possible planning outcomes via a \textbf{box-and-whisker plot}\footnote{A \textit{box-and-whisker plot} is a statistical representation commonly used to depict the range of observations present in a data set. The line inside the box indicates the median of the data, while the ends of the box indicate the upper (75\%) and lower (25\%) quartiles. The whiskers may extend to the highest and lowest observations, though statistical outliers are often depicted as individual points beyond the whiskers. Notably, a box-and-whisker plot visually shows the full range of observations without averaging and without presuming that observations follow any kind of parametric distribution.}. An example is shown in Figure \ref{fig:gcmbxplt} for projected temperatures by decade in San Francisco (we note that this example is a climate variable, not a planning outcome). The strength of this visual depiction lies in its intuitive representation of the range and distribution of projected outcomes. Once this range is presented, recommendations from different planning models can be synthesized to make decisions based on the severity of outcomes that could occur if planning decisions are made in accord with a climate projection that proves to be wrong.
    \item Review every category of planning decisions and determine the \textbf{appropriate risk threshold}. For example, planning to the median may be appropriate in cases where the risks are not severe (e.g., projected PV panel output). In other cases, planning to extremes will be more appropriate (e.g., water availability for cooling power plants). Decision-makers may opt to set different thresholds (e.g., 50th, 75th, or 95th percentile) for different types of planning decisions, depending on risk tolerance.
    \item In cases where the implications of unmitigated risks are extreme, more thorough analysis will be necessary to ensure that investments are commensurate with the magnitude of the risks, and that society is prepared to shoulder the implications of whatever trade-offs are ultimately made (see Recommendation 8).
\end{enumerate}
This exercise will produce planning recommendations for each climate scenario, while assessing the potential risks (see Recommendation 4) associated with planning to one risk threshold versus another.
% The outcome of this exercise should be a list of planning decisions (including a range of their possible outcomes as informed by the different climate scenarios), accompanied by risk thresholds set by decision-makers that can then be used by stakeholders for infrastructure planning.

 % gcmbxplt

%Research tasks: (a) determine how a climate variable might impact a given planning decision, e.g., where in the box-and-whisker plot does it make sense to look and (b) which variables contribute more or less to which planning decisions (sensitivity analysis)
\textbf{\textsc{In the near term,}} infrastructure planning models should be run with the full range of climate projections specified in Recommendation 1. Box-and-whisker plots of planning outcomes should be generated and a sensitivity analysis of different climate impacts on infrastructure planning model assumptions should be performed. \textbf{\textsc{Ultimately,}} the range of planning outcomes should be obtained via updated infrastructure planning models (see Recommendation 6). The risk analysis and appropriate risk tolerances should be revisited and applied to future decisions.

%\pagebreak
\subsection{Grappling with uncertainty and its consequences}
\label{sec:recommendations:uncertainty}

The complex interactions between climate and electricity infrastructure leave room for significant uncertainty. 
Modelers will likely make mistakes and overlook some climate interactions.
Should this be the case, processes must be in place that facilitate iteration and allow us to refine grid planning models as new information comes to light. 

\subsubsection*{\textbf{\textsc{Recommendation 8:} Assess the magnitude and severity of evolving risk scenarios; re-evaluate risk tolerances in light of evolving risks and/or allocate resources to respond effectively.}}

\paragraph{\textbf{\textsc{Barrier}}} 
Infrastructure planning decisions have always balanced investments in performance against the probability and implications of failure. 
Today, these decisions are made primarily based on quantitative risk metrics informed by historic data and/or expert judgment \cite{sed2016}.
As climate trends change the nature and severity of viable risk scenarios, the current approach makes us vulnerable to systematic errors in our understanding of current and future threats.
These errors prevent us from taking effective and targeted actions to mitigate evolving risks; we are shouldering the consequences without fully understanding what they are.

\paragraph{\textbf{\textsc{Solution}}} Recommendation 4 cites the need for climate-aware risk evaluation protocols, while Recommendation 7 proposes that risk tolerances be set.
Here, we recommend detailed analysis of unmitigated risk scenarios to ensure that diverse stakeholder groups are prepared for the consequences.
It may be necessary to re-evaluate risk thresholds should it be determined that potential consequences are too high.
\textbf{\textsc{In the near term,}} risk assessment efforts should focus on ensuring that the likelihood and ramifications of unmitigated risk scenarios are fully understood.
CPUC and other decision-makers should engage with local communities, emergency response agencies, and disaster relief funds (among others) to ensure that the diverse groups who will shoulder the burden of real-time response are adequately prepared to do so.
\textbf{\textsc{Ultimately,}} efforts to identify and quantify climate-related risks, and decisions about how to allocate limited resources must be coordinated across stakeholder groups who are positioned to mitigate risks, and those who will need to respond.
These efforts can aid more efficient and effective responses to risk scenarios that are realized, and provide a basis for reaching consensus about the magnitude of investment appropriate for mitigating risk scenarios that carry societal consequences more severe than those experienced to date.

\subsubsection*{\textbf{\textsc{Recommendation 9:} Planning agencies need to think critically about how infrastructure needs could change due to climate impacts on population and behavior.}}

\paragraph{\textbf{\textsc{Barrier}}} 
There remains a great deal of uncertainty around how people will respond to climate change.
%We have already begun to see examples of climate mitigation and adaptation measures changing grid infrastructure needs.
We may anticipate some trends, but high uncertainty around how or if to respond may mean it is too soon for preemptive action.
Several examples have emerged.
Expansion of the wildland-urban interface has created new load centers and increased the implications of wildfire ignition events \cite{strikeforce2019}.
Public safety power shutoffs have been used to mitigate severe wildfire risks, but outages pose safety risks to vulnerable populations during extreme heat waves.
Service interruptions during wildfire events may impact the ability of electric vehicle owners to evacuate affected areas.

There are other examples we may anticipate but have not yet confronted.
For example, researchers are studying potential population displacement due to climate change \cite{Scientists2018}.
However, little discussion has focused on infrastructure needs to support displaced communities \cite{Sherbinin2011}. Without proper planning, temporary or permanent displacement of populations---within, to, and from California---could lead to considerable strain on infrastructure systems.

\paragraph{\textbf{\textsc{Solution}}} 
The high degree of uncertainty may make it untenable to prioritize these risks over more immediate needs, but we should still prepare to take action.
Stakeholder engagement efforts should focus on enumerating the wide range of risk scenarios that could unfold as people respond and adapt to climate change, and evaluate the implications on grid infrastructure and operations. 
To the extent possible, we should anticipate and plan for these types of risk scenarios and timely, coordinated, and effective responses.

% CONCLUSIONS AND POLICY IMPLICATIONS
\section{Conclusions and Policy Implications}
\label{sec:conclusions}
Climate change will fundamentally alter the operating conditions and risk exposure of electric grid infrastructure.
An advanced body of research has developed climate data to anticipate these risks and inform long-term planning decisions.
Additional research may be needed to characterize certain climate-grid interactions.
Where data do exist, substantial research has been done to examine how climate trends will impact grid assets, for example due to warming and sea level rise.

The current work examines the technical details of climate-grid interactions, opportunities and barriers to use climate information to inform long-term investments in grid infrastructure.
We detail nine recommendations that provide guidance to coordinating entities (for example regulators or policymakers) positioned to enable and advance climate awareness in grid planning processes in California.
These recommendations are grounded in existing grid infrastructure planning processes, and are informed by best-practices established in other sectors.

We find that much of the data necessary to support climate-aware decision-making are readily available, including projections of temperature, precipitation, snowpack, and wildfire risk.
However, climate projections characterizing changes in the severity of high wind-speed events---which can both physically damage grid infrastructure and exacerbate wildfire risks---are needed. 
Grid planning agencies rely heavily on load forecasts which need to be revised to reflect how climate trends and mitigation efforts will alter load.

Regulators and policymakers will play a critical role in disseminating climate data, and in advancing infrastructure planning models to incorporate these data.
An important first step is to ensure that decision-makers throughout the state have access to and use climate and load projections as inputs to grid infrastructure planning models.
Clear guidance should be issued regarding which data are appropriate to use for different planning decisions.
Research funding is needed to generate datasets that are lacking, and to study grid-climate interactions that are not well understood.

Though there is a great deal of uncertainty associated with making long-term decisions in the face of climate change, that need not prevent decision-makers from mitigating recognized risks.
Existing regulatory processes are well-suited for regulators to detail prescriptive guidance about how to make infrastructure investment decisions in light of existing uncertainties.
Transparent risk thresholds informed by current knowledge will ensure that planning decisions align with evolving societal needs and priorities, and allow greater insight into unmitigated risks.
Planning thresholds will need to be revisited as our understanding of evolving risks advances.
Developing a more detailed understanding of the risks that climate-grid interactions could pose can support regulatory agencies in determining how resources are allocated, and whether rate increases will be necessary to cover the costs of mitigating untenable risks.

\section*{Acknowledgements}

Many individuals provided helpful guidance and discussions. From the California Public Utilities Commission, we are grateful to Rachel Peterson, Suzanne Casazza, Reese Rogers, Patrick Young, and Sonya Ziaja for their expert insights and thoughtful input. We also benefited from the wisdom of Guido Franco from the California Energy Commission, Mohit Chhabra from the Natural Resources Defense Council, John Holmes and Ben Wender from the National Academy of Sciences, Evan Mills from Lawrence Berkeley National Laboratory, Jessie Knapstein from Pacific Gas \& Electric, Peter Alstone from Humboldt State University, and Kripa Jagannathan, Julia Szinai, Ian Bolliger, and Adam Orford from UC Berkeley. Also at UC Berkeley, our research advisers, Duncan Callaway and Scott Moura, and Steven Weissman from the Goldman School of Public Policy (who provided the impetus for this project) contributed invaluable feedback and insights throughout. Any remaining errors and/or oversights are solely our own.

A.B. was supported by the National Science Foundation Graduate Research Fellowship under Grant No. DGE 1752814. L.D. was supported by a 2019 Seed Fund Award from CITRIS and the Banatao Institute at the University of California.

%%TC:ignore
\section*{Acronyms}

\noindent\begin{tabulary}{0.49\linewidth}{l L}
AC & Air Conditioning \\
CAISO & California Independent Systems Operator \\
CEC & California Energy Commission \\
CMIP & Coupled Model Intercomparison Project \\
CPUC & California Public Utilities Commission \\
DER & Distributed Energy Resource \\
%E3 & Energy + Environmental Economics \\
EV & Electric Vehicle \\
FERC & Federal Energy Regulatory Commission \\
GCM & Global Circulation Model \\
GRC & General Rate Case \\
IEPR & Integrated Energy Policy Report \\
IPCC & Intergovernmental Panel on Climate Change \\
IPR & Integrated Resource Plan[ning] 
\end{tabulary} 
\begin{tabulary}{0.51\linewidth}{l L}
LOCA & Localized Constructed Analogs \\
LSE & Load-Serving Entity \\
O\&M & Operations and Maintenance \\
OIR & Order Instituting Rulemaking \\
PV & Photovoltaic \\
RAMP & Risk Assessment Mitigation Phase [proceeding] \\
RCP & Representative Concentration Pathway \\
RESOLVE & Renewable Energy Solutions Model \\
RPS & Renewable Portfolio Standard \\
RSP & Reference System Plan \\
SERVM & Strategic Energy \& Risk Valuation Model \\
S-MAP & Safety Model Assessment Proceeding \\
TPP & Transmission Planning Process \\
\end{tabulary}
%%TC:endignore

%% The Appendices part is started with the command \appendix;
%% appendix sections are then done as normal sections
%%TC:ignore
\appendix
%\pagebreak
% \input{appendices.tex}

\section{Electricity infrastructure planning in California}
\label{sec:appCAplanning}

Building on Section \ref{sec:currentplanning}, we return here to a more detailed discussion of electricity system planning in California. We focus on selected infrastructure planning activities and current approaches to modeling and data streams at the three state entities that share responsibility for energy- and electricity-related planning and oversight \cite{CEC2015CAenergyagencies}. 

The California Energy Commission (CEC) conducts a broad suite of activities related to energy policy and planning. CEC's responsibilities include developing integrated policy strategies for the state, funding research and demonstration projects, approving sites for large thermal power plants, setting efficiency standards for buildings and appliances, and certifying renewable energy resources for compliance with the state's Renewable Portfolio Standard. 
The California Public Utilities Commission (CPUC) regulates essential infrastructure and utility services. CPUC has jurisdictional oversight over investor-owned public utilities that provide  electricity and natural gas service, and over private telecommunications, water, and transportation companies. Within the electric utility sector, CPUC is responsible for authorizing procurement that is in the public interest, regulating rates, and ensuring safety.\footnote{Notably, other entities such as Community Choice Aggregation (CCA) programs and Electric Service Providers (ESPs) also serve customer loads in California but CPUC jurisdiction over them is limited. We exclude them from further consideration here.} 
The California Independent Systems Operator (CAISO) operates electric transmission infrastructure and oversees wholesale electric power markets within its planning region, which covers most of the state of California.

\subsection{Load forecasting}
\label{sec:currentplanning:load}

The CEC issues official load forecasts in conjunction with its Integrated Energy Policy Report (IEPR), which is released every two years with updates in the intervening years \cite{iepr2017, CEC2018IEPRupdate}. 
Forecasts are generated for \textit{high}, \textit{mid}, and \textit{low} energy demand cases which are designed with varying assumptions about economic growth, electricity and gas prices, energy efficiency, PV generation and EV adoption, weather conditions, and climate change impacts. 

\begin{figure}[t]
\centering
\includegraphics[width=1\linewidth]{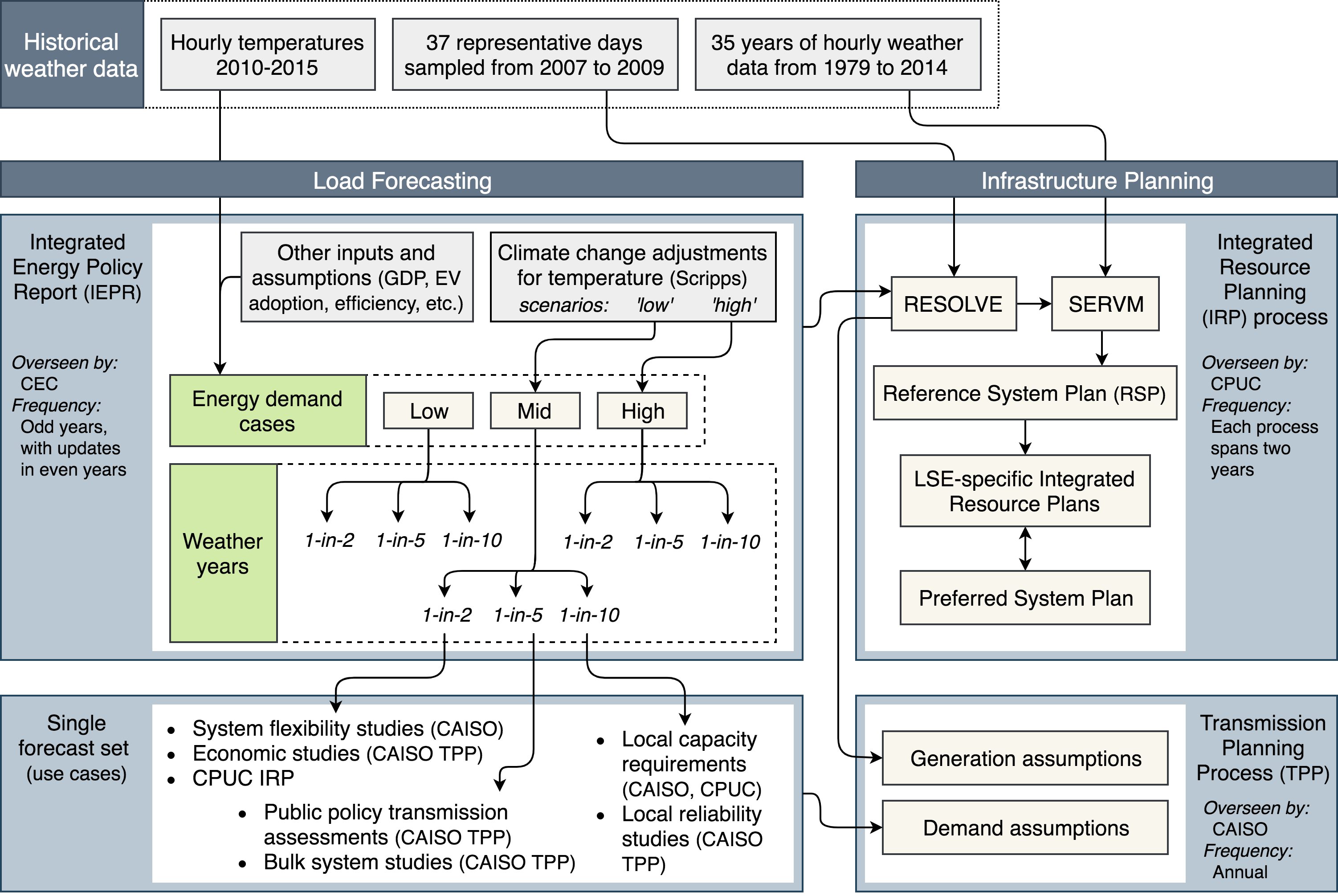}
\caption{Data flows in existing California planning processes pertaining to electricity infrastructure planning.}
\label{fig:CAdataflows}
\end{figure}

%The 2017 IEPR demand forecasts include scenarios related to energy efficiency, PV generation, and changing end consumption patterns. They also include 
To date, the underlying weather inputs into CEC's demand forecasts have been based on a sampling of historical weather data. Hourly temperature data collected over fifteen years from 2000 to 2015 was used to inform the demand forecasts published with the 2017 IEPR. This historical data was subjected to a random sampling process to identify representative distributions of hourly temperatures. These distributions enabled an estimation of hourly peak demand as well as the ratio of demand in other hours to the peak demand hour \cite[p.14-28]{CEC2017IEPRdatainputstranscript}. Multiple `weather years'---full years of hourly temperature data that follow distributions representative of the historical record---were generated for each energy demand case by repeating this process (Figure \ref{fig:CAdataflows}). From these multiple weather years, CEC issued load forecasts for 1-in-2, 1-in-5, and 1-in-10 weather years intended to represent, for example, a high-demand (hot) year encountered on average once every 10 years \cite{CPUC2016IRPassumptions,CaliforniaPublicUtilitiesCommission2019}. These scenarios are meant to account for uncertainty and contingencies related to extreme weather. However, each forecast is still based on the historical record, which limits its applicability for planning to future incidences of extreme values.

Climate change adjustments to historical weather trends are based on temperature scenarios developed by the Scripps Institution of Oceanography \cite[p.176]{iepr2017}. 
The \textit{low} demand case incorporates no additional impacts from climate change, while the \textit{mid} and \textit{high} demand cases use adjustments based on `low' and `high' scenarios of temperature increases \cite{iepr2017} (Figure \ref{fig:CAdataflows}). 
%The broader research community cannot comment further on the appropriateness of these scenarios, as 
Documentation does not specify which specific Scripps temperature scenarios were chosen or which assumptions are contained therein (see \cite{Orford2019} for a more thorough discussion). These temperature adjustments are used to estimate the additional energy consumption and peak impacts for residential and commercial customers within specified planning zones. 
However, the methods currently used to account for climate impacts on projected demand are not comprehensive: only temperature (specifically, cooling and heating degree days and annual maximum) is considered in the climate change adjustments to historical data used in the 2017 IEPR. 
Further research is needed to comprehensively account for these impacts.

In addition to the forecast scenarios discussed above---the \textit{low}, \textit{mid}, and \textit{high} energy demand cases, each with their own 1-in-2, 1-in-5, and 1-in-10 weather years---the CEC publishes a \textit{single forecast set} intended for use in statewide planning processes at the CPUC and CAISO. This common forecast uses the \textit{mid} energy demand case discussed above, with its weather years assigned to specific planning purposes (Figure \ref{fig:CAdataflows}) \cite{iepr2017,CEC2018IEPRupdate}.\footnote{The \textit{single forecast set} also incorporates scenarios for additional energy efficiency savings and PV adoption. We omit further discussion of these here due to our primary focus on weather and climate data used in the demand forecasts.}

\subsection{Integrated Resource Planning (IRP)}
\label{sec:currentplanning:IRP}
The CPUC's IRP process is the state's primary venue for long-term planning and procurement decision-making related to electricity infrastructure \cite{CPUC2019IRPweb}. 
The biennial process begins with two phases of system-wide modeling activities that aim to identify a portfolio of resources that meet policy goals \cite{CPUC2018IRPreqsforLSEs}. 
First, the RESOLVE capacity expansion model (see Section \ref{sec:infrastructure} for an overview of infrastructure planning model types) from Energy + Environmental Economics (E3) \cite{E32019RESOLVEweb} is used to develop large-scale planning and procurement scenarios that guide investment decisions. 
RESOLVE uses the official demand forecasts from CEC's IEPR process as inputs. 
The model aims to identify the optimal transmission and distribution investments that meet demand while satisfying reliability and policy constraints within a particular geographic area. 
RESOLVE also relies on historical weather data: the model simulates system operations for 37 days ``sampled from the historical meteorological record of the period 2007-2009" that are weighted to ``produce a reasonable representation of complete distributions of potential conditions" \cite[pp.49-50]{E32017RESOLVEdocumentation}. 

Since RESOLVE models the state's CAISO planning area as a single node \cite{E32017RESOLVEdocumentation},\footnote{RESOLVE models electricity transmission in the western U.S. with six nodes. Four correspond to California balancing authorities, with CAISO modeled as a single node. The other three CA zones correspond to the Los Angeles Department of Water and Power (LADWP), the Imperial Irrigation District (IID), and, together, the Balancing Authority of Northern California (BANC) and Turlock Irrigation District (TID). The other two nodes ``represent regional aggregations of out-of-state balancing authorities" in the northwest and southwest \cite[p.48]{E32017RESOLVEdocumentation}.} its results require post-processing to provide insight about more granular spatial considerations, such as the best way to meet demand in particular load pockets. 
A second model, the Strategic Energy \& Risk Valuation Model (SERVM) from Astrap\'{e} Consulting \cite{Astrape2019SERVMweb}, is used by the CPUC to determine whether a full year of hourly demand can be met into the future without adverse reliability impacts with the planning scenario proposed by RESOLVE. 
SERVM tests RESOLVE outputs with a probabilistic distribution of weather years. Within SERVM, the original (unaveraged) historic weather values are preserved and scaled up to annual and peak values projected in future load forecasts \cite{CaliforniaPublicUtilitiesCommission2019}. 
Historical weather data is used to account for a variety of weather conditions that are crucial inputs to predicting future electricity generation from solar and wind resources (e.g., cloud cover and temperature in addition to wind speed and solar irradiance) and capture resource variability \cite{CaliforniaPublicUtilitiesCommission2019}. 
SERVM achieves greater spatial granularity within the CAISO region by dividing California into eight modeling areas, but it is still a simplification of a full network model \cite{CaliforniaPublicUtilitiesCommission2019}.\footnote{SERVM divides California into eight modeling regions. The four non-CAISO areas are IID, LADWP, BANC, and TID. The CAISO region is divided into four areas corresponding to the utility territories of San Diego Gas \& Electric, Pacific Gas \& Electric (divided in two), and, together, Southern California Edison and Valley Electric Association \cite[p.16]{CaliforniaPublicUtilitiesCommission2019}.} 

%For centralized resource procurement, the CEC takes proposed renewable resource capacity portfolios and allocates this capacity to transmission-level substations controlled by CAISO for planning purposes \cite{CEC2019IRPallocation}. 

The results of this system-wide modeling process are adopted as the Reference System Plan (RSP) \cite{CPUC2018IRPreqsforLSEs}.
Load serving entities (LSEs), which include regulated utilities, maintain control of specific decisions related to infrastructure planning within their systems. 
LSEs use the RSP as a guide to develop their preferred approaches to meet planning requirements within their own Integrated Resource Plans. 
LSE-specific IRPs are submitted to the CPUC, which compiles them into a new system-wide portfolio, evaluates them for compliance, and authorizes procurement based on the amount of capacity deemed necessary for reliability. as part of the General Rate Case (GRC), which occurs every three years \cite{CPUC2019IRPweb,CPUC2018IRPreqsforLSEs}. (Additional iterative steps may occur between individual LSE plans and the statewide portfolio prior to approval.) In developing their procurement plans, some LSEs currently use a 1-in-10 historical weather year to prepare for climate-related risks \cite{Rogers2019,Billinton2015b}.

\subsection{Transmission Planning Process (TPP)}
\label{sec:currentplanning:TPP}

Each year, CAISO conducts the state's transmission planning process to determine system needs. The process aims to ``identify potential system limitations as well as opportunities for system reinforcements that improve reliability and efficiency" \cite{CAISO2019TPPweb}. Finalized in March 2018, the data inputs for the 2019 TPP cycle are designed to be consistent with the CPUC's IRP \cite{CAISO2018TPPassumptions}. Specifically, the assumptions pertaining to energy generation in the state come from the CPUC's RESOLVE model, while the demand forecasts come from the CEC's IEPR (Figure \ref{fig:CAdataflows}).

\subsection{Risk assessment}
\label{sec:currentplanning:risk}

In 2013, the CPUC issued the so-called ``Risk OIR'' to initiate a new paradigm for increasing transparency into how risks are evaluated and prioritized \cite{risk_rulemaking}.
These analyses inform investment plans that are documented in GRC filings and subsequently acted upon.
The Risk OIR gave rise to two new proceedings: the Safety Model Assessment Proceeding (S-MAP) and the Risk Assessment Mitigation Phase (RAMP) proceeding.
%These proceedings were instigated to enhance transparency into methods for assessing and prioritizing risk mitigation measures proposed by utilities.

S-MAP is intended to provide documentation requisite for both experts and non-experts to understand the logical processes, input variables, and quantitative methods utilities use to examine risk exposure \cite{risk_decision}.
In the context of these proceedings, the concept of ``risk'' encompasses anything that poses a safety threat (e.g., wildfire, employee safety, public safety).
Risk metrics include both reported safety incidents (e.g., ``overhead wires down'', ``fire ignitions'', ``employee serious injuries or fatalities'') and preventative actions that were taken (e.g., employee training, tree trimming, and equipment inspection) \cite{risk_metrics}.

Initial guidelines for RAMP were to enumerate ``the top ten asset-related risks for which the utility expects to seek recovery in the GRC'' focusing on ``asset conditions and mitigating risks to those assets'' \cite{smap_decision}.
These guidelines, however, are intended merely as a starting point; there is a stated expectation that the contents of RAMP filings will evolve as risk assessment protocols become more mature.
The S-MAP and RAMP proceedings are designed to facilitate this process by providing transparency into existing risk evaluate practices, and by subjecting them to scrutiny (for example, see \cite{sed2016, sed2018}).
Though risk assessment methods are still evolving, the criticism (and praise) regarding these two proceedings shows that demonstrable progress has already been made.
Ultimately, the intention is for the S-MAP and RAMP proceedings to tend towards a risk evaluation paradigm that is consistent across all utilities regulated by the CPUC.

\subsection{Additional activities}
\label{sec:currentplanning:additional}

While S-MAP and RAMP are designed to address general risk assessment protocols, a number of other proceedings address additional risk assessment needs on a more targeted ad-hoc basis.
Examples include submission of utility wildfire mitigation plans \cite{wildfire_oir}, efforts to map wildfire threat \cite{firemap_oir}, and to assess physical threats of grid infrastructure \cite{security_oir}.
Other proceedings (e.g., Distribution Resources Planning \cite{CPUC2018DRPweb}) may also impact electricity infrastructure planning more broadly by prompting regulated utilities and other stakeholders to consider additional priorities such as distributed renewables integration, public health and safety, and overall system performance.

Notably, recent events related to the increasing frequency and magnitude of wildfires in California have prompted increasing discussion about de-energizing electricity lines and other infrastructure to reduce risk in high-fire conditions. 
A move towards more frequent de-energization as an operating principle may, in turn, prompt additional focus on distributed operation and planning in contrast to the centralized system planning approach we describe above. 
This could ultimately lead to a qualitatively different planning framework that more effectively accounts for trade-offs and co-benefits between decisions made in alignment with diverse planning objectives that are currently represented in different proceedings.

\section{Mechanisms for power production}
\label{sec:appGen}

Here, we provide additional detail on technology-specific first principles for electricity generators.

\subsection{Solar photovoltaics}
\label{sec:appGen:PV}

Power generation from PV cells depends on the available solar irradiance and surrounding temperature, both of which vary with site location \cite{Dubey2013}. 
The sun's rays provide photons that act as a current source and trigger the flow of electrons within the PV cell. Solar irradiance is therefore directly proportional to the  current within the cell, and a PV cell operating in half-sun will produce roughly half the power of a cell in full sun \cite[Ch.5.4]{Masters2013}. On the other hand, higher temperatures affect the voltage within the PV cell by speeding up electron-hole recombination before electrons can generate electricity that leaves the cell. PV cell efficiency is typically reported under standard test conditions, which are defined as solar irradiance of 1 $kW/m^2$ (1 sun), cell temperature of 25\degree C, and air mass ratio of 1.5 (AM1.5)\footnote{The air mass ratio is less critical to the current discussion and is included here primarily for completeness. It describes the amount of air (i.e., atmosphere) that sunlight must travel through to reach the earth. AM1.5 is the standard value used for mid-latitudes, including the contiguous U.S.} \cite[Ch.5.6]{Masters2013}. The decrease in maximum power generation with temperature varies by PV technology and manufacturer, but can be around 0.24-0.45\% per degree Celsius \cite[Ch.5.7]{Masters2013}. Notably, the operating temperature of a PV cell is also affected by the solar irradiance incident on the cell. The following equation is used to estimate this impact:

\begin{equation}
    T_{cell} = T_{amb} + \left(\frac{NOCT - 20 \degree C}{0.8 kW/m^2}\right) \times S
\end{equation}

\noindent where $T_{cell}$ (\degree C) is the operating cell temperature, $T_{amb}$ (\degree C) is the actual ambient air temperature, $S$ is actual solar insolation $kW/m^2$, and $NOCT$ is a standard cell-specific parameter provided by the manufacturer that corresponds to the expected cell temperature under 20\degree C ambient temperature, 0.8 $kW/m^2$ solar irradiation, and 1 m/s wind speed. A PV panel with NOCT 46 operating at 30\degree C (86\degree F) and 1-sun irradiance will therefore have an internal cell temperature of 62.5\degree C (145\degree F) and deliver 17\% less electricity than is indicated by its rated maximum power capacity \cite[Ch.5.7]{Masters2013}. Changes from today's performance due to climate change impacts on temperature and solar irradiance can be estimated in a similar manner.

\subsection{Wind}
\label{sec:appGen:wind}
Regional wind speed patterns play a critical role in decisions related to siting, design, and operations of wind farms.
The relationship is best explained by examining the equation relating wind power production to wind speed \cite{dunn2019}:
\begin{equation}
\label{eq:wind_power}
    P = k \cdot \min\{v, v_r\}^3
\end{equation}
where $P$ is power produced, $v$ is the current wind speed, $v_r$ is the rated wind speed of the turbine, and $k$ is a lumped parameter describing mechanical and aerodynamic properties of a particular wind turbine design.
Equation \ref{eq:wind_power} holds up to some cut-out wind speed (often 25 mph) which is always greater than $v_r$.
Turbines are turned off when conditions exceed the cut-off wind speed to prevent damage.

Because $P$ scales in proportion to wind speed cubed, wind power production is sensitive to wind speed conditions at a specific site.
Because $v_r$ changes for different turbine designs, design decisions are made based on the statistical properties of the wind field at a particular site.
Therefore, changes in wind field characteristics could reduce power production from existing wind farms.

\subsection{Natural gas and other thermal power generation}
\label{sec:appGen:naturalgas}

A typical thermoelectric power plant burns fuel to create high-pressure steam. 
The steam turns turbine blades, thereby powering an electric generator and converting heat to electricity. 
The waste heat is released to a low-temperature sink. The thermal efficiency of this cycle can be described as:
\begin{equation}
\label{eq:naturalgasthermo}
    \eta = \frac{T_{high} - T_{low}}{T_{high}} = \frac{W_{net}}{T_{high}}
\end{equation}
where \(\eta\) corresponds to the thermal efficiency of the plant, \(T_{high}\) is the high temperature achieved by burning fuel, \(T_{low}\) is the temperature of the low-temperature sink, and \(W_{net}\) is the work done by the temperature differential.

A thermal plant's electricity production potential thus depends on the temperature differential between the hot steam from the combustion process and the low-temperature sink of waste heat. 
This low-temperature sink, typically a nearby body of water or the surrounding air, is crucial to cooling the plant, and its ambient temperature directly affects plant operation and efficiency.  
Thermal plants throughout the U.S. rely on once-through or recirculating water cooling or dry cooling \cite{Peer2016, Miara2017, Lee2018}. Water-cooled systems are also subject to thermal discharge limits \cite{McCall2016}.

Natural gas plants are the most common types of thermal power plants in California and are frequently used to compensate for variability in solar and wind generation. However, similar considerations apply to other types of thermal power generation, including concentrating solar power plants and some biomass plants.

%\section{Table of recommendations}

%%TC:endignore
%\pagebreak
%% References
%% Following citation commands can be used in the body text:
%% Usage of \cite is as follows:
%%   \cite{key}          ==>>  [#]
%%   \cite[chap. 2]{key} ==>>  [#, chap. 2]
%%   \citet{key}         ==>>  Author [#]

%% References with bibTeX database:
\bibliographystyle{model1-num-names}
\bibliography{Projects-ClimateAdaptation.bib,gspp_project.bib}

%% Authors are advised to submit their bibtex database files. They are requested to list a bibtex style file in the manuscript if they do not want to use model1-num-names.bst.

%% References without bibTeX database:
% \begin{thebibliography}{00}
%% \bibitem must have the following form:
%%   \bibitem{key}...
%%
% \bibitem{}
% \end{thebibliography}

%%TC:ignore
%\detailtexcount{elsarticle-template-1-num}
%%TC:endignore

\end{document}